\documentclass[fleqn,usenatbib]{mnras}

\usepackage{newtxtext,newtxmath}

\usepackage[T1]{fontenc}

\DeclareRobustCommand{\VAN}[3]{#2}
\let\VANthebibliography\thebibliography
\def\thebibliography{\DeclareRobustCommand{\VAN}[3]{##3}\VANthebibliography}

\usepackage{graphicx}	
\usepackage{amsmath}	

\newcommand{\Omegac}{\Omega_{\rm c}}
\newcommand{\Ic}{I_{\rm c}}
\newcommand{\Nc}{N_{\rm c}}
\newcommand{\xic}{\xi_{\rm c}}
\newcommand{\tauc}{\tau_{\rm c}}
\newcommand{\sigmac}{\sigma_{\rm c}}

\newcommand{\Omegas}{\Omega_{\rm s}}
\newcommand{\Is}{I_{\rm s}}
\newcommand{\Ns}{N_{\rm s}}
\newcommand{\xis}{\xi_{\rm s}}
\newcommand{\taus}{\tau_{\rm s}}
\newcommand{\sigmas}{\sigma_{\rm s}}

\title[Analysing radio pulsar timing noise]{Analysing radio pulsar timing noise with a Kalman filter: a demonstration involving PSR J1359$-$6038}

\author[N. J. O'Neill et al.]{
Nicholas J. O'Neill,$^{1,2}$\thanks{E-mail: noneill@student.unimelb.edu.au}
Patrick M. Meyers,$^{3,1,2}$
Andrew Melatos$^{1,2}$
\\
$^{1}$School of Physics, University of Melbourne, Parkville, VIC 3010, Australia\\
$^{2}$Australian Research Council Centre of Excellence for Gravitational Wave Discovery (OzGrav), University of Melbourne, Parkville, VIC 3010, Australia\\
$^{3}$Theoretical Astrophysics Group, California Institute of Technology, Pasadena, CA 91125, USA
}

\date{Accepted XXX. Received YYY; in original form ZZZ}

\pubyear{2023}

\begin{document}
\label{firstpage}
\pagerange{\pageref{firstpage}--\pageref{lastpage}}
\maketitle

\begin{abstract}
In the standard two-component crust-superfluid model of a neutron star, timing noise can arise when the two components are perturbed by stochastic torques. Here it is demonstrated how to analyse fluctuations in radio pulse times of arrival with a Kalman filter to measure physical properties of the two-component model, including the crust-superfluid coupling time-scale and the variances of the crust and superfluid torques. The analysis technique, validated previously on synthetic data, is applied to observations with the Molonglo Observatory Synthesis Telescope of the representative pulsar PSR J1359$-$6038. It is shown that the two-component model is preferred to a one-component model, with log Bayes factor $6.81 \pm 0.02$. The coupling time-scale and the torque variances on the crust and superfluid are measured with $90\%$ confidence to be $10^{7.1^{+0.8}_{-0.5}}$ $\rm{s}$ and $10^{-24.0^{+0.4}_{-5.6}}$ $\rm{rad^2~s^{-3}}$ and $10^{-21.7^{+3.5}_{-0.9}}$ $\rm{rad^2~s^{-3}}$ respectively.
\end{abstract}




\section{Introduction}

High-precision pulsar timing data are a rich and plentiful resource for probing the properties of neutron star interiors. They show that pulsars decelerate secularly with small, random fluctuations away from the secular trend in the form of occasional glitches \citep{1996MNRAS.282..677S, 2010MNRAS.402.1027H, 2011MNRAS.414.1679E, 2015IJMPD..2430008H} and persistent pulsar timing noise \citep{Shannon_2010, 2016MNRAS.458.2161L, 2019MNRAS.489.3810P, Lower2020}. The random fluctuations can be related to internal physics, such as far-from-equilibrium processes involving superfluid vortices \citep{1975Natur.256...25A, 2011MNRAS.415.1611W}, type II superconductor flux tubes \citep{1998ApJ...492..267R, drummond:2017wfg}, and hydrodynamic turbulence \citep{1970Natur.227..791G, Melatos2014}. One prominent structural theory supported by these observations is the two-component model, in which neutron stars are composed of a rigid outer crust and a superfluid core \citep{Baym_1969}. The crust and superfluid are coupled by forces, which act to restore corotation and therefore damp observable fluctuations in the crust's angular velocity, as happens during post-glitch recoveries. With respect to timing noise specifically, the tendency to restore corotation through crust-superfluid coupling should be detectable statistically, e.g. through the auto-correlation time-scale  \citep{Price_2012}. One advantage is that timing noise occurs all the time, so there is an abundance of data to work with.

The two-component model of neutron stars was originally motivated by observations of pulsar glitches \citep{Baym_1969}. Glitches are rare events in which a pulsar spins up impulsively. Following the event, the spin-down rate (which sometimes also changes during the event) returns partially or wholly to the rate before the event, due to some friction-like interaction between the crust and superfluid operating over some time-scale $\tau$ \citep{1975Natur.256...25A, 1978ApJ...224..969L, 1985Natur.316...27P, 2015IJMPD..2430008H, 2022atcc.book..219A, 2022Univ....8..641Z, 2022RPPh...85l6901A}. As timing noise also involves a perturbation of the crust's angular velocity, it is likely to either drive or be driven by glitch-like differential rotation between the crust and superfluid, which is damped by the same restoring forces that bring the two components back into corotation after a glitch \citep{1978ApJ...225..582L, 2010MNRAS.409.1253V, Price_2012, 2017MNRAS.471.4827G}. We aim to measure the coupling between the two components by looking for statistical evidence of its associated relaxation time-scale $\tau$ in timing noise data.

Many previous timing noise studies focus on calculating ensemble averaged quantities derived from the whole data set such as the noise amplitude \citep{Boynton_1972, Groth_1975_III, Cordes_1980, 1980ApJ...239..640C, Shannon_2010} and power spectrum \citep{Van_Haasteren_2009, 2010MNRAS.402.1027H, 2016MNRAS.458.2161L, 2020MNRAS.497.3264G, 2019MNRAS.489.3810P, 2020MNRAS.494.2012P}. Although ensemble averages are good for measuring some quantities, they do not make use of the information contained in the specific, time-ordered sequence of times of arrival, that is, the specific realization of the random process presented to the observer \citep{2023MNRAS.522.4880V}. In this paper, we seek to extract this extra information using a Kalman filter, a tool commonly used for signal processing in electrical engineering applications \citep{kalman1960}. Given a model for how a system evolves with time (here the two-component model), the Kalman filter tracks the specific random fluctuations in the observed time series and estimates the most likely state of the system at every time step, i.e. the most likely values of the dynamical variables (the crust and superfluid angular velocities) in the two-component model. Moreover, when combined with a nested sampler, the Kalman filter estimates the most likely values of the static parameters in the model, e.g. the crust-superfluid coupling coefficient and the variances of the stochastic torques. 
Two previous papers \citep{Meyers2021a, Meyers2021b} explained this approach and validated it with synthetic data. 

In this paper we demonstrate in principle that the approach works successfully on real data by applying it to Molonglo Observatory Synthesis Telescope observations of PSR J1359$-$6038 \citep{2017PASA...34...45B, Lower2020}. This object is chosen, because its timing noise power spectrum is similar to that predicted by the simplest version of the two-component model, as discussed below (see Section \ref{sec:data} and Appendix \ref{app:analytic derivation of power spectrum}). It also offers relatively good quality data, with $429$ times of arrival (TOAs) over $3.5$ years of monitoring, with an average TOA uncertainty of $5.1 \times 10^{-5}\rm{s}$. Other pulsars can be analysed but an analysis of a sample of objects is outside the scope of this paper, whose goal is to demonstrate the real-world applicability of the method by way of a worked example, to pave the way for fuller studies in the future. 

The paper is structured as follows. In Section \ref{sec:two-component model}, the two-component model is introduced. In Section \ref{sec:kalman tracking and estimation}, the Kalman filter and its implementation are explained, including the question of parameter identifiability and the choice of priors based on existing observations. In Section \ref{sec:data} we discuss the properties of the data from PSR J1359$-$6038, including the number of observations and measurement uncertainties, and describe the conversion from TOAs to local frequencies. In Section \ref{sec:results} we present the parameter estimates for PSR J1359$-$6038. In Section \ref{sec:one-component model} we present a Bayesian comparison between the one-component and two-component models. In Section \ref{sec:conclusion}, we conclude by interpreting the results and validating them with Monte Carlo posterior checks to verify their statistical significance. 
The appendices contain for completeness an analytic derivation of the power spectral density of the two-component model, the explicit formulas for the matrices used in the Kalman filter, the Kalman filter recurrence relations, calculations demonstrating the accuracy of the parameter estimates as a function of data volume and measurement errors, and details of the Bayesian comparison of the two-component model with a simpler, one-component model.

\section{Two-component model}
\label{sec:two-component model}

The two-component model of a neutron star consists of a rigid crust and superfluid core, which are assumed to rotate uniformly, with angular velocities $\Omega_{\rm c}$ and $\Omega_{\rm s}$ respectively. The components obey the equations of motion \citep{Baym_1969, 2017MNRAS.471.4827G}
\begin{align}
    \Ic\frac{\textrm{d}\Omegac}{\textrm{d}t} &= - \frac{\Ic}{\tauc}\left(\Omegac - \Omegas\right) + \Nc + \xic(t) \label{eq:crust de},\\
    \Is\frac{\textrm{d}\Omegas}{\textrm{d}t} &= - \frac{\Is}{\taus}\left(\Omegas - \Omegac\right) + \Ns + \xis(t) \label{eq:core de},
\end{align}
where the subscripts ``c'' and ``s'' label the crust and the superfluid respectively, $\Ic$ and $\Is$ are the moments of inertia, $\Nc$ and $\Ns$ are constant external torques, $\xic$ and $\xis$ are stochastic torques, and $\tauc$ and $\taus$ are coupling time-scales. The astrophysical origins of the torques on the right-hand sides of (\ref{eq:crust de}) and (\ref{eq:core de}) are discussed below.
The stochastic torques obey Gaussian, white noise statistics with
\begin{align}
\langle \xi_{\rm c,s}(t)\rangle &= 0\\
\label{eq:white_noise_torque_covariance}\langle \xi_{\rm c,s}(t)\xi_{\rm c,s}(t')\rangle &=\sigma_{\rm c,s}^2\delta(t-t'),
\end{align}
where $\langle\ldots\rangle$ denotes the ensemble average, and $\sigmac$ and $\sigmas$ are noise amplitudes.

The coupling between the crust and superfluid is assumed to act like friction between the two components, reducing their relative velocity. In this paper $|\Omega_{\textrm{c}} - \Omega_{\textrm{s}}|$ is assumed to be small enough that the coupling torque is approximately linear in this difference on the right hand sides of equations (\ref{eq:crust de}) and (\ref{eq:core de}).
Over a typical observing time-scale (decades), and over the relaxation time-scales $\tau_{\rm c}$ and $\tau_{\rm s}$ (typically weeks) \citep{Price_2012}, the torques $N_{\rm c}$ and $N_{\rm s}$ can be approximated as constants. $N_{\rm c}$ represents the net secular external torque applied to the crust, including  the electromagnetic radiation reaction torque \citep{goldreich:1969wiu} and the gravitational radiation reaction torque \citep{1969ApJ...157.1395O, 1969ApJ...158L..71F}. $N_{\rm s}$ represents the net secular torque on the superfluid which includes electromagnetic and gravitational components, if the superfluid is threaded by a corotating magnetic field and possesses a time-varying mass or current quadrupole moment respectively.
The torque $\xi_{\rm{c}}$ represents the net random torque acting on the crust, which may be negligible in a nonaccreting radio pulsar, except perhaps in the presence of seismic activity \citep{2006ApJ...652.1531M, 2010MNRAS.407L..54C, 2022MNRAS.514.1628K, 2022MNRAS.511.3365G}. $\xi_{\rm{s}}$ represents the net random torque acting on the superfluid, including the torque from superfluid vortex unpinning \citep{1975Natur.256...25A, 2011MNRAS.415.1611W}. The stochastic torques drive timing noise in the observable $\Omega_{\rm c}$, even in the scenario $\xi_{\rm{c}} = 0$ and $\xi_{\rm{s}} \neq 0$.

Equations (\ref{eq:crust de}) and (\ref{eq:core de}) have been generalized recently to multiple internal components by \citet{2023MNRAS.520.2813A}. We restrict the analysis in this paper to two components, because the available data are insufficient to constrain a more complicated model. Indeed, the data are insufficient even to constrain (\ref{eq:crust de}) and (\ref{eq:core de}) fully, as discussed in Section \ref{sec:kalman tracking and estimation}.

\section{Kalman tracking and estimation}
\label{sec:kalman tracking and estimation}

Given an observed time series $\Omega_{\rm c}(t_1), \dots , \Omega_{\textrm{c}}(t_N)$, one can use a Kalman filter to infer the most probable sequence of states $\Omega_{\rm c}(t_i)$ and $\Omega_{\rm s}(t_i)$ $(1\leq i \leq N)$ traversed by the system described by (\ref{eq:crust de}) and (\ref{eq:core de}). In this paper, the Kalman filter assumes the measurements are the true state of the system plus some Gaussian measurement noise. The Kalman filter then uses the assumption that the system evolves according to (\ref{eq:crust de}) and (\ref{eq:core de}) to separate out the real evolution of the system from the measurement noise on a most probable basis, which gives estimates of $\Omega_{\rm c}(t_i)$ and $\Omega_{\rm s}(t_i)$ at each $t_i$ together with an error on each estimate. This procedure, termed Kalman tracking, is described in Section \ref{ssec:kalman tracking}. The Kalman filter's ability to recover the model parameters from $\Omega_{\rm c}$ data is assessed in Section \ref{ssec:identifiability}. To estimate the parameters, one calculates the Kalman filter likelihood (of the model producing the observed data) as a function of the static parameters to find their most probable values given $\Omega_{\rm c}(t_i)$. This procedure, termed Kalman estimation, is described in Section \ref{ssec:kalman filter likelihood}. The results depend on the priors, which are specified in Section \ref{ssec:prior distribution} together with their astrophysical motivations.

\subsection{Kalman tracking}
\label{ssec:kalman tracking}

The state of the system at time $t_i$ is denoted by 
\begin{align}
\boldsymbol X_i = 
\begin{pmatrix}
\Omega_{\rm c}(t_i)\\
\Omega_{\rm s}(t_i)
\end{pmatrix}. \label{eq:state definition}
\end{align}
The Kalman filter predicts the next state of the system given the current one.
Solving (\ref{eq:crust de}) and (\ref{eq:core de}) gives a discrete equation for updating the state from time $t_{i-1}$ to $t_i$,
\begin{align}
\boldsymbol X_i = \boldsymbol F_i \boldsymbol X_{i-1} + \boldsymbol T_i + \boldsymbol w_i. \label{eq:kalman update}
\end{align}
Equation (\ref{eq:kalman update}) multiples $\boldsymbol X_{i-1}$ by a transition matrix $\boldsymbol F_i$, which comes from the coupling terms in (\ref{eq:crust de}) and (\ref{eq:core de}); adds a vector $\boldsymbol T_i$, which comes from the deterministic torques; and adds a random vector $\boldsymbol w_i$, which comes from the stochastic torques and is drawn from a Gaussian distribution with 
\begin{align}
\label{eq:process noise mean}
\langle \boldsymbol w_i \rangle &= \boldsymbol 0\\
\label{eq:process noise cov}
\langle \boldsymbol w_i \boldsymbol w_j \rangle &= \boldsymbol Q_i \delta_{ij}.
\end{align}
Explicit expressions for $\boldsymbol F_i$, $\boldsymbol T_i$ and $\boldsymbol Q_i$ are given in Appendix \ref{app:full state space representation}.

For general physical systems, the state and measurements at $t_i$ may be related in a complicated non-linear way. In this paper, however, the relation is simple: the measured and true $\Omega_{\rm c}(t_i)$ differ by an additive measurement error, and $\Omega_{\rm s}(t_i)$ is hidden, i.e. it cannot be measured directly at all. 
Mathematically, we encode this by saying that a measurement $\boldsymbol Y_i$ at time $t_i$ is related to the state of the system $\boldsymbol X_i$ by
\begin{align}
\label{eq:kalman measurement}
\boldsymbol Y_i = \boldsymbol C \boldsymbol X_i + \boldsymbol v_i.
\end{align}
In (\ref{eq:kalman measurement}), $\boldsymbol v_i$ is the measurement noise and is drawn from a normal distribution with 
\begin{align}
\label{eq:measurement noise}
\langle \boldsymbol v_i \rangle &= \boldsymbol 0\\
\langle \boldsymbol v_i \boldsymbol v_j\rangle &= \boldsymbol R_i \delta_{ij}.
\end{align}
$\boldsymbol C$ is the observation matrix which determines which components of the state can be measured. In radio pulsar timing experiments, only the crust can be measured, since the core is hidden from view, implying \footnote{In the future, gravitational waves
radiated from the core may be detectable, whereupon $\Omega_{\rm s}(t_i)$ could be measured directly.}
\begin{align}
\label{eq:observation matrix}
\boldsymbol C = (1,0).
\end{align}

The Kalman filter computes the expected evolution of the system through (\ref{eq:kalman update}). It updates the expected evolution with new information from the measurement at $t_i$ through (\ref{eq:kalman measurement}), separating the process noise ${\boldsymbol w}_i$ from the measurement noise ${\boldsymbol v}_i$. The detailed implementation of this two-step, predictor-corrector algorithm is discussed in Appendix \ref{app:explanation of the Kalman filter} to help the interested reader reproduce the results in Sections \ref{sec:results} and \ref{sec:one-component model}.

\subsection{Identifiability}
\label{ssec:identifiability}

In general, there is no guarantee that all six of the static parameters $\tau_{\rm c}, \tau_{\rm s}, \sigma_{\rm c}, \sigma_{\rm s}, N_{\rm c}, N_{\rm s}$ in (\ref{eq:crust de}) and (\ref{eq:core de}) can be identified from the time series $\Omega_{\rm c}(t_1), \dots, \Omega_{\rm c}(t_N)$, even if the data volume is arbitrarily large ($N\rightarrow \infty$). Certain parameters cannot be separated from the rest and can be estimated only in combination. This issue, known as {\em identifiability}, is widely recognized in electrical engineering applications. A number of formal techniques have been developed to handle it, as summarized by ~\citet{BELLMAN1970329} for example. In this section, we analyze what combinations of the static parameters in (\ref{eq:crust de}) and (\ref{eq:core de}) are identifiable, in order to interpret the posterior distribution.

As a starting point, it is instructive to express (\ref{eq:crust de}) and (\ref{eq:core de}) as an equation for $\Omega_{\rm c}$ on its own, viz.
\begin{align}
\label{eq:em_only_omgddot}
\ddot\Omega_{\rm c} &= -\left(\frac{1}{\tauc} + \frac{1}{\taus} \right)\dot\Omega_{\rm c} + \left(\frac{\Nc}{\taus\Ic} + \frac{\Ns}{\tauc\Is}\right) + \frac{\xi_{\rm{c}}}{\taus\Ic} + \frac{\xi_{\rm{s}}}{\tauc\Is} + \frac{\dot \xi_{\rm{c}}}{I_{\rm c}}.
\end{align}
Equation (\ref{eq:em_only_omgddot}) determines $\Omega_{\rm c}(t)$ fully, supplemented by initial conditions $\Omega_{\rm c}(t_1)$ and $\dot{\Omega}_{\rm c}(t_1)$. Therefore the static parameter combinations that appear in (\ref{eq:em_only_omgddot}) are identifiable --- that is, they can be estimated from the data --- but they cannot be decomposed into their elements. For example, $\tau_{\rm c}$ and $\tau_{\rm s}$ cannot be estimated separately, because they appear in the irreducible combination $\tau_{\rm c}^{-1} + \tau_{\rm s}^{-1}$ in the first term on the right-hand side of (\ref{eq:em_only_omgddot}), and no additional, independent equation of motion exists beyond (\ref{eq:em_only_omgddot}), in which $\tau_{\rm c}$ and $\tau_{\rm s}$ appear separately.

The first term on the right-hand side of (\ref{eq:em_only_omgddot}) contains the parameter
\begin{align}
\label{eq:reduced_relaxation_time}
\tau &= \frac{\tauc\taus}{\tauc+\taus},
\end{align}
which is a combined relaxation time-scale for the system as a whole.
The second term (when multiplied by $\tau$) contains the parameter
\begin{align}
\label{eq:ensemble_averaged_spindown}\langle \dot\Omega_{\rm c}\rangle &= \frac{1}{\tauc+\taus}\left(\tau_{\rm c} \frac{\Nc}{\Ic}+\taus \frac{\Ns}{\Is}\right),
\end{align}
which is the ensemble-averaged frequency derivative of the pulsar.
The noiseless evolution of $\Omega_{\rm c}$ is governed solely by $\tau$ and $\langle \dot \Omega_{\rm c} \rangle$ so these parameters are easiest to estimate. For notational convenience, and to assist with physical interpretation, we also introduce the complementary parameter combinations,
\begin{align}
\label{eq:time_scale_ratio}
r &= \frac{\taus}{\tauc}\\
\label{eq:lag} \langle \Omega_{\rm c} - \Omega_{\rm s} \rangle &= \tau \left(\frac{\Nc}{\Ic}-\frac{\Ns}{\Is}\right).
\end{align}
In (\ref{eq:time_scale_ratio}) and (\ref{eq:lag}), $r$ is the ratio of relaxation time-scales, which equals $I_{\rm s}/I_{\rm c}$ when the coupling torques form an action-reaction pair, and $\langle \Omega_{\rm c} - \Omega_{\rm s} \rangle$ is the ensemble-averaged angular velocity lag between the crust and superfluid. 

It is less straightforward to read off by sight whether the noise amplitudes $Q_{\rm c} = \sigma_{\rm{c}}^2/I_{\rm c}^2$ and $Q_{\rm s} = \sigma_{\rm{s}}^2/I_{\rm s}^2$ can be estimated by the Kalman filter. However, in Appendix \ref{app:parameter estimation on simulated frequencies}, the question is answered empirically using synthetic data. It turns out that the Kalman filter can estimate $Q_{\rm c}$ reliably (in addition to $\tau$ and $\langle \dot{\Omega}_{\rm c} \rangle$) but usually not $Q_{\rm s}$. An analytic treatment of the identifiability of $Q_{\rm c}$ and $Q_{\rm s}$ is given in Appendix \ref{app:identifiability of Qc, Qs and R}.
The analysis suggests that it is easiest to recover the parameter combinations $r, \tau, Q_{\rm c}, Q_{\rm s}, \langle \Omega_{\rm c} - \Omega_{\rm s} \rangle, \langle \dot \Omega_{\rm c} \rangle$, and we switch to them in what follows.

\subsection{Kalman filter likelihood}
\label{ssec:kalman filter likelihood}

We construct a likelihood function $p(\boldsymbol Y_{1:N} | \boldsymbol \theta)$ with $\boldsymbol Y_{1:N} = \{\boldsymbol Y_1, \boldsymbol Y_2, ...,\boldsymbol Y_{N}$\} for the data given a choice of parameters. We calculate $p(\boldsymbol Y_{1:N} | \boldsymbol \theta)$ from the optimal state sequence output by the Kalman filter. Intuitively, if the parameters are chosen well (i.e. near their true values), the model's trajectory in time matches the data closely, and $p(\boldsymbol Y_{1:N}|\boldsymbol \theta)$ is relatively large.

Let $\boldsymbol \epsilon_i$ be the difference between the measurement and the prediction of the state at $t_i$, and let $\boldsymbol S_i$ be the covariance of $\boldsymbol \epsilon_i$. Then Bayes's rule gives the posterior distribution of $\boldsymbol \theta$ as
\begin{align}
\nonumber \ln p(\boldsymbol \theta \vert \boldsymbol Y_{1:N}) = &- \frac{1}{2} \sum_{i=1}^{N-1} \left[ N_{\boldsymbol Y} \ln(2\pi) + \ln \det (\boldsymbol{S}_{i}) + \boldsymbol{\epsilon}_{i}^T \boldsymbol{S}_{i}^{-1} \boldsymbol{\epsilon}_{i} \right]\\
&+ \ln p(\boldsymbol \theta) - \ln(Z), \label{eq:kalman_likelihood}
\end{align}
where $N_{\boldsymbol Y}$ is the dimension of $\boldsymbol Y_i$, which in our case is always unity, $p(\boldsymbol \theta)$ is the prior distribution of $\boldsymbol \theta$, and $Z$ is the evidence, which acts as a normalisation constant. Appendix \ref{app:explanation of the Kalman filter} supplies more details on the derivation of (\ref{eq:kalman_likelihood}). The \texttt{dynesty} sampler~\citep{speagle2020oas} is used to efficiently sample the posterior distribution.

\subsection{Prior distribution}
\label{ssec:prior distribution}

In the absence of other information, we choose log-uniform priors for the parameters that are positive definite and plausibly span several decades, namely $r, \tau, Q_{\rm c}$ and $Q_{\rm s}$. We choose uniform priors for the parameters with small ranges, namely $\langle \dot \Omega_{\rm c} \rangle$ and $\langle \Omega_{\rm c} - \Omega_{\rm s} \rangle$. As more pulsars are analysed in the future, and a picture of the distribution of (say) $\tau$ across the population emerges, it may become appropriate to use more informative priors in future work.

We can use previous measurements of glitch relaxation time-scales to infer reasonable bounds on $\tau$, motivated by (\ref{eq:reduced_relaxation_time}). The data in \citet{Yu_2013} imply $10^5 \leq \tau / (1 \, \rm{s}) \leq 10^8$. Models of stellar structure suggest $10^{-2} \leq r = \tau_{\rm s}/\tau_{\rm c} = I_{\rm s}/I_{\rm c} \leq 10^2$ \citep{link:1999iek,lyne:2000cie,2011MNRAS.414.1679E,chamel:2012cwl}. Reasonable bounds on $Q_{\rm c}$ and $Q_{\rm s}$ follow from previous, independent measurements of timing noise amplitude. For the crust one measures typically $10^{-30} \leq Q_{\rm c} / (1 \, {\rm rad^2 \, s^{-3}}) \leq 10^{-16}$ \citep{1985ApJS...59..343C, Cerri_Serim_2019, Lower2020, 2023MNRAS.522.4880V}.
In the absence of additional information, we assume the same range for $Q_{\rm s}$, noting that radio pulsar timing experiments cannot track the core to measure $Q_{\rm s}$ directly. 
For $\langle \dot \Omega_{\rm c} \rangle$, a central estimate $\langle \dot \Omega_{\rm c} \rangle_{\rm{cent}}$ and an uncertainty $\sigma$ can be obtained by fitting a linear trend to the time series $\Omega_{\rm c}(t_i)$. We adopt a uniform prior for $\langle \dot \Omega_{\rm c} \rangle$ with a range of $[\langle \dot \Omega_{\rm c} \rangle_{\rm{cent}} - 1000\sigma, \langle \dot \Omega_{\rm c} \rangle_{\rm{cent}} + 1000\sigma]$. For the crust-core lag, one extreme case $N_{\rm c}/I_{\rm c} \ll N_{\rm s}/I_{\rm s} \leq 0$ implies $\langle \Omega_{\rm c} - \Omega_{\rm s} \rangle \sim \tau \langle \dot \Omega_{\rm c} \rangle$. The opposite extreme $N_{\rm s}/I_{\rm s} \ll N_{\rm c}/I_{\rm c} \leq 0$ implies $\langle \Omega_{\rm c} - \Omega_{\rm s} \rangle \sim - \tau \langle \dot \Omega_{\rm c} \rangle$. 
We therefore assume $\tau_{\rm{max}} \langle \dot \Omega_{\rm c} \rangle_{\rm{min}} \leq \langle \Omega_{\rm c} - \Omega_{\rm s} \rangle \leq -\tau_{\rm{max}} \langle \dot \Omega_{\rm c} \rangle_{\rm{min}}$, where the subscripts "max" and "min" denote the maximum and minimum values of the prior range given earlier in this section. For most pulsars, one finds $-10^{-11} \leq \langle \dot \Omega_{\rm c} \rangle \leq 0$ \citep{Helfand_1980} and hence $-10^{-3} \leq \langle \Omega_{\rm c} - \Omega_{\rm s} \rangle \leq 10^{-3}$. For more details on justifying the prior distribution, the reader is referred to \citet{Meyers2021a, Meyers2021b}.

\section{Data}
\label{sec:data}

The data used in this paper come from the UTMOST pulsar observing program\footnote{\url{https://github.com/Molonglo/TimingDataRelease1/}} \citep{Lower2020}. They are in the form of barycentered pulse times of arrival (TOAs). To assist parameter estimation, we select a test object with a relatively large number of TOAs with relatively small error bars. We also avoid objects exhibiting glitches, because glitches are not included in the dynamical model (\ref{eq:crust de}) and (\ref{eq:core de}). An object which satisfies these criteria is PSR J1359$-$6038. The UTMOST data release for PSR J1359$-$6038 contains 429 TOAs, one of which we discard as an outlier \footnote{The TOA at MJD 58190.7 differs from the trend of its neighbouring TOAs by $\approx 8$ times more than its error bars. An investigation in \citet{2022MNRAS.512.1469D} suggests that the outlier was due to observatory conditions since a similar offset occurs in data for other pulsars at the same observatory at that time.}. The average TOA error is $51~\mu\rm{s}$. The TOAs span 1263 days. In \citet{Lower2020} the timing noise in PSR J1359$-$6038 is measured to have a spectral index of $\gamma = -5.1^{+0.8}_{-0.9}$, i.e. the power spectral density of the phase residuals scales $\propto f^{\gamma}$ at $f \gtrsim 10^{-8}~{\rm Hz}$. This is comparable to the theoretical prediction for the two-component model (\ref{eq:crust de}) and (\ref{eq:core de}), of $\gamma = -4$, given in (\ref{eq:spectrum_limit}).

The Kalman filter in Section \ref{sec:kalman tracking and estimation} ingests pulse frequency data. Hence the phase information in the TOAs must be converted to local frequencies $\Omega_{\rm c}(t_1)$, $\dots$, $\Omega_{\rm c}(t_N)$. This is done using the standard pulsar timing software package, {\sc tempo2} \citep{Edwards_2006a, Edwards_2006b}. 
To generate a local frequency estimate, we feed a set, $S_{n}$, of consecutive TOAs into {\sc tempo2} and extract $\Omega(\overline{t}_n)$ and its associated error, where $\overline{t}_n$ is the average of the TOAs in the set $S_n$. We construct the disjoint sets, $S_1, S_2, ..., S_N$, from the TOAs, $t_1, t_2, ..., t_M$ ($M \geq N$), starting with $t_1$, according to the following two rules: (i) $S_n$ contains $t_i$ and all $t_j > t_i$ with $|t_j - t_i| < 10\, {\rm days}$; and (ii) there must be at least three TOAs per set, and the 10-day window is lengthened when necessary to achieve this.

The fitting process yields $62$ frequency data points spanning $1220$ days (the $\overline{t}_n$ values span a slightly shorter time than the $t_i$ values). We find $\Omega_{\rm c}(t_i) = \Omega_{\rm c}(t_1) + (t_i - t_1) \dot{\Omega}_{\rm c}(t_1)$ to an excellent approximation, with $\Omega_{\rm c}(t_1) = 49.2767 ~\rm{rad}~\rm{s}^{-1}$ and $\dot{\Omega}_{\rm c}(t_1) = -2.4472 \times 10^{-12}~\rm{rad}~\rm{s}^{-2}$. Subtracting the linear, long-term trend yields the frequency residuals plotted in Fig. \ref{fig:freqs_J1359-6038}. The standard deviation of the residuals is $1.7 \times 10^{-8}~\rm{rad}~\rm{s}^{-1}$. The mean uncertainty is $7.8 \times 10^{-9}~\rm{rad}~\rm{s}^{-1}$. However the size of the error bars varies with the errors of the original TOAs, the number of TOAs used in a fit, and the time span of a fit. The interval from MJD 57400 to MJD 57900 features TOAs with relatively large error bars and spacings; the median uncertainty on $\Omega(t_i)$ therein is $1.3 \times 10^{-8} ~\rm{rad}~\rm{s}^{-1}$ compared to $2.6 \times 10^{-9} ~\rm{rad}~\rm{s}^{-1}$ for the measurements outside that interval. 
Some fitted frequencies deviate significantly from the linear trend, possibly due to an issue with fitting to TOAs with large errors. We remove two outliers with residuals from the trend below  $-10^{-7}~\rm{rad}~\rm{s}^{-1}$.

The conversion from TOAs to frequencies is imperfect. Fitting frequencies filters the timing noise on the time-scale over which the fit is done, so some information is lost. 
The effect of filtering is tested on simulated TOAs in Appendix \ref{app:tempo2 fits to local frequencies}. We find that it causes $Q_{\rm{c}}$ to be slightly underestimated and makes $\tau$ harder to recover.

\section{Estimated parameters of the two-component model}
\label{sec:results}

In this section, we apply the Kalman tracking and estimation procedure described in Section \ref{sec:kalman tracking and estimation} to the UTMOST data described in Section \ref{sec:data}. Marginalized posterior distributions for the static parameters that can be identified reliably, namely $\tau$, $Q_{\rm{c}}$, $Q_{\rm{s}}$ and $\langle \dot \Omega_{\rm{c}} \rangle$, are presented in Section \ref{ssec:joint posterior distribution}. The four parameters are discussed individually in greater detail in Sections \ref{ssec:coupling time-scale}--\ref{ssec:average spin-down rate} and interpreted astrophysically. The two parameters that cannot be identified reliably, namely $r$ and $\langle \Omega_{\rm{c}} - \Omega_{\rm{s}} \rangle$, are discussed in Section \ref{ssec:unidentifiable parameters}.

\subsection{Joint posterior distribution}
\label{ssec:joint posterior distribution}

\begin{figure*}
    \centering
    \includegraphics[width=\textwidth]{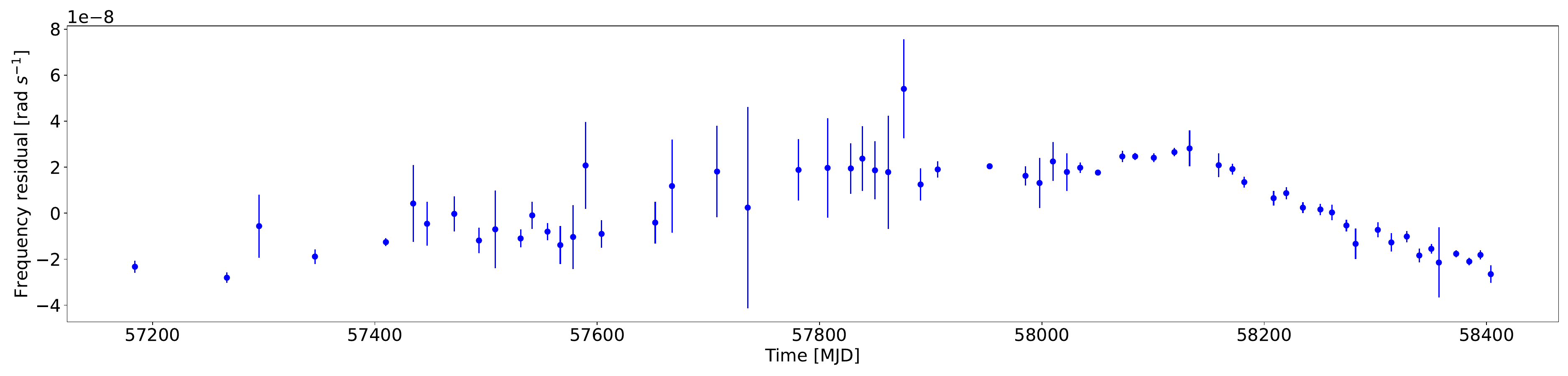}
    \caption{Residuals of frequencies fitted to 428 TOAs versus time (units: MJD) for the PSR J1359$-$6038 after subtracting a linear trend ($\Omega(t)[\textrm{rad}~\textrm{s}^{-1}] = 49.2888 - 2.4472 \times 10^{-12} t[\rm{s}]$) and removing two outliers due to poor fitting. The vertical scale is in units of $10^{-8} ~\rm{rad ~s}^{-1}$. The frequencies and error bars are calculated using {\sc tempo2} from the TOAs and their uncertainties in the UTMOST data release.}
    \label{fig:freqs_J1359-6038}
\end{figure*}

The six-dimensional joint posterior distribution calculated from the PSR J1359$-$6038 frequency data in Fig. \ref{fig:freqs_J1359-6038} using equation (\ref{eq:kalman_likelihood}) is displayed in the traditional format of a corner plot in Appendix \ref{app:full posterior distribution for PSR J1359-6038}. Four parameters can be estimated reliably and exhibit well-formed peaks: $\tau$, $Q_{\rm{c}}$, $Q_{\rm{s}}$ and $\langle \dot \Omega_{\rm{c}} \rangle$. Two parameters, $r$ and $\langle \Omega_{\rm{c}} - \Omega_{\rm{s}} \rangle$, cannot be estimated reliably and rail against the prior bounds.

\begin{figure*}
    \centering
    \includegraphics[width=\textwidth]{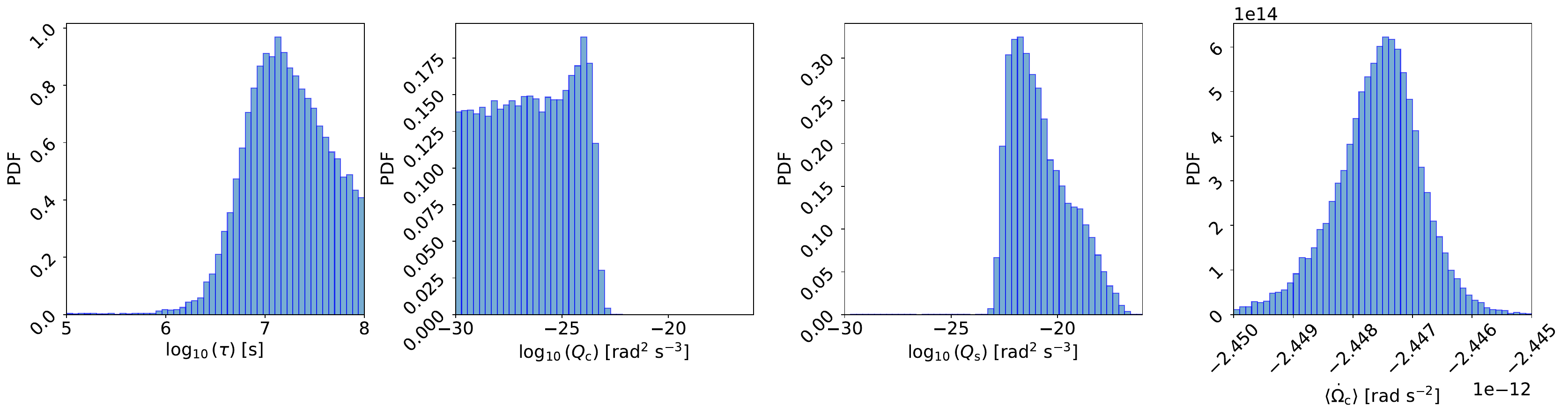}
    \caption{Marginalized posterior distributions for the four parameters, $\tau$, $Q_{\rm{c}} = \sigma_{\rm{c}}^2/I_{\rm{c}}^2$, $Q_{\rm{s}} = \sigma_{\rm{s}}^2/I_{\rm{s}}^2$ and $\langle \dot \Omega_{\rm{c}} \rangle$ that can be estimated reliably for the two-component model applied to PSR J1359$-$6038. The marginalized posteriors are plotted as histograms spanning the prior range, except for $\langle \dot \Omega_{\rm{c}} \rangle$ where only the region around the peak is shown. The units of the parameters are specified on the horizontal axis of each panel. The vertical axis carries a linear scale, showing the value of the marginalised PDF for that parameter.}
    \label{fig:marginalised_posteriors_J1359-6038}
\end{figure*}

One-dimensional cross-sections of the posterior, marginalized over all but one parameter, are displayed as histograms in Fig. \ref{fig:marginalised_posteriors_J1359-6038} for the four parameters that can be estimated reliably. The marginalized posteriors in Fig. \ref{fig:marginalised_posteriors_J1359-6038} are unimodal, with the peaks falling comfortably within the respective prior ranges in Table \ref{tab:prior_table}. There are distinct peaks for $\tau$, $Q_{\rm{c}}$, $Q_{\rm{s}}$ and $\langle \dot \Omega_{\rm{c}} \rangle$. They occur at $\tau = 1.3 \times 10^7 ~\rm{s}$, $Q_{\rm c} = 1.05 \times 10^{-24} ~\rm{rad^2~s^{-3}}$, $Q_{\rm s} = 1.82 \times 10^{-22} ~\rm{rad^2~s^{-3}}$ and $\langle \dot \Omega_{\rm{c}} \rangle = -2.4475 \times 10^{-12} ~\rm{rad~s^{-2}}$. 

The widths of the $90\%$ confidence intervals of the histograms in Fig. \ref{fig:marginalised_posteriors_J1359-6038} are $1.3$ dex, $6.0$ dex, $4.4$ dex and $2.4 \times 10^{-15} ~\rm{rad~s^{-2}}$ for $\tau, Q_{\rm{c}}, Q_{\rm{s}}, \langle \dot \Omega_{\rm{c}} \rangle$ respectively. The $90\%$ confidence intervals cover $43\%, 43\%, 31\%$, and $1.4\%$ of their respective prior ranges, confirming that the four parameters are identified reliably.

Numerical values for the parameter estimates for PSR J1359$-$6038 conditional on the two-component model given by (\ref{eq:crust de}) and (\ref{eq:core de}) are summarized for completeness in Table \ref{tab:twocomp_result_table}.

\bgroup
\def\arraystretch{1.5}
\begin{table}
\centering
\caption{Prior distributions introduced in Section \ref{ssec:prior distribution} for the parameter estimation carried out in Figs. \ref{fig:posteriors_J1359-6038} and \ref{fig:posteriors_J1359-6038_onecomp}. The rightmost column indicates the prior distribution, $p(\cdot)$, on each parameter. $\mathcal{U}(a,b)$ indicates a uniform distribution between $a$ and $b$. The priors of the first four (last two) parameters are log-uniform (uniform).}
\begin{tabular}{c|l|l|l}
\hline
Parameter    & Units               & Prior\\
\hline
$r$    &                     & $\log \mathcal U(10^{-2}, 10^2)$\\
$\tau$ & $\rm{s}$       & $\log \mathcal U(10^{5}, 10^{8})$\\
$Q_{\rm{c}}$  & $\rm{rad^2~s^{-3}}$ & $\log \mathcal{U}(10^{-30}, 10^{-16})$\\
$Q_{\rm{s}}$  & $\rm{rad^2~s^{-3}}$ & $\log \mathcal{U}(10^{-30}, 10^{-16})$\\
$\langle\Omega_{\rm c}-\Omega_{\rm s}\rangle$ & $\rm{rad~s^{-1}}$ & $\mathcal{U}(-10^{-3}, 10^{-3})$\\    
$\langle \dot\Omega_{\rm c}\rangle$ & $\rm{rad~s^{-2}}$ & $\mathcal{U}(-2.5323 \times 10^{-12}, -2.3621 \times 10^{-12})$\\
\hline
\end{tabular}
\label{tab:prior_table}
\end{table}
\egroup

\bgroup
\def\arraystretch{1.5}
\begin{table*}
    \centering
    \caption{Static parameters inferred by the Kalman tracker and nested sampler for the two-component model, extracted from the six-dimensional joint posterior in Fig. \ref{fig:posteriors_J1359-6038}. The $\tau$, $Q_{\rm{c}}$, $Q_{\rm{s}}$ and $\langle \dot \Omega_{\rm{c}} \rangle$ rows correspond to the four identifiable parameters, whose marginalized posteriors are plotted in Fig. \ref{fig:marginalised_posteriors_J1359-6038}. The $r$ and $\langle \Omega_{\rm{c}} - \Omega_{\rm{s}} \rangle$ rows correspond to the unidentifiable parameters discussed in Section \ref{ssec:unidentifiable parameters}. The fourth and fifth columns list two complementary measures of the dispersion in the posterior, viz.\ the full-width half-maximum (FWHM) and 90\% confidence intervals respectively.}
\begin{tabular}{c|c|c|c|c}
\hline
Parameter    & Units               & Peak value  & FWHM interval& $90\%$ confidence interval\\
\hline
$\log_{10} r$    &                     & ~~$-0.6$    & $(-2.0, 2.0)$ & $(-1.8, 1.8)$\\
$\log_{10} \tau$ & $\rm{s}$       & ~~~~$7.1$   & $(6.7, 7.9)$ & $(6.6, 7.9)$\\
$\log_{10} Q_{\rm{c}}$  & $\rm{rad^2~s^{-3}}$ & $-24.0$     & $(-30.0, -23.3)$ & $(-29.6, -23.6)$\\
$\log_{10} Q_{\rm{s}}$  & $\rm{rad^2~s^{-3}}$ & $-21.7$     & $(-22.7, -19.9)$ & $(-22.6, -18.2)$\\
$\langle \Omega_{\rm c} - \Omega_{\rm s} \rangle$ & $\rm{rad~s^{-1}}$ & ~~~$3 \times 10^{-4}$ & $(-1.0 \times 10^{-3}, 1.0 \times 10^{-3})$ & $(-9.0 \times 10^{-4}$, $9.0 \times 10^{-4})$\\    
$\langle \dot \Omega_{\rm c} \rangle$ & $\rm{rad~s^{-2}}$ & ~~$-2.4475 \times 10^{-12}$ & $(-2.4482 \times 10^{-12}, -2.4468 \times 10^{-12})$ & $(-2.4489 \times 10^{-12}, -2.4465 \times 10^{-12})$\\
\hline
\end{tabular}
\label{tab:twocomp_result_table}
\end{table*}
\egroup

\subsection{Coupling time-scale}
\label{ssec:coupling time-scale}

In Fig. \ref{fig:marginalised_posteriors_J1359-6038} we find that the marginalised posterior for $\tau$ has a well defined peak. With 90\% confidence we recover $3.8 \times 10^{6} \leq \tau/(1\,{\rm s}) \leq 7.6 \times 10^7$.
This result is broadly consistent with previous measurements of timing noise and post-glitch relaxation time-scales.
\citet{Price_2012} computed the auto-correlation time-scale of timing residuals in the objects PSR B1133+16 and PSR B1933+16 and obtained $\approx 10 \, {\rm days}$ and $\approx 20 \, {\rm days}$ respectively. Post-glitch relaxation time-scales have been measured previously by many authors \citep{1990Natur.346..822M, 1993ApJ...409..345A, 1996MNRAS.282..677S, lyne:2000cie, 2000MNRAS.317..843W, 2001ApJ...548..447W, 2002ASPC..271..357D, 2010MNRAS.404..289Y, 2010MNRAS.409.1253V, 2011MNRAS.414.1679E, Yu_2013,  2021A&A...647A..25E, 2021MNRAS.508.3251L, 2022MNRAS.511..425G}. They mostly range from $\approx 10 \, {\rm days}$ to $\approx 3\times 10^2 \, {\rm days}$ and occasionally reach as high as $\approx 10^3 \, {\rm days}$. The 90\% confidence interval in Fig. \ref{fig:marginalised_posteriors_J1359-6038} overlaps this range.
The prior on $\tau$ in Table \ref{tab:prior_table} is motivated by glitch observations, so the overlap is expected, but it is significant that the posterior peaks comfortably within the prior range.

\subsection{Torque noise amplitudes}
\label{ssec:torque_noise_amplitudes}

The marginalised posteriors for $Q_{\rm{c}}$ and $Q_{\rm{s}}$ are both unimodal. With 90\% confidence we recover $\log_{10} (Q_{\rm{c}}/ 1\, {\rm{rad^2~s^{-3}}}) = -24.0^{+0.4}_{-5.6}$ and $\log_{10} (Q_{\rm{s}}/ 1\, {\rm{rad^2~s^{-3}}}) = -21.7^{+3.5}_{-0.9}$. 

In previous timing noise analyses, the statistics of the timing residuals are modelled by a power spectral density (PSD) of the form \citep{2016MNRAS.458.2161L, 2019MNRAS.489.3810P, Lower2020}
\begin{align}
P(f) = \frac{A^2}{12\pi^2 f_{\rm{yr}}^{3}} \left(\frac{f}{f_{\rm{yr}}}\right)^{\gamma},
\label{eq:gw_spectrum1}
\end{align}
where $A^2$ is a dimensionless squared amplitude, $\gamma$ is a dimensionless exponent, and we define the reference frequency $f_{\rm yr} = (1 \, {\rm yr})^{-1}$.

For $f \gg \tau^{-1} \sim 10^{-7} \, {\rm Hz}$, the two-component model (\ref{eq:crust de})--(\ref{eq:white_noise_torque_covariance}) is consistent with (\ref{eq:gw_spectrum1}), as demonstrated analytically in Appendix \ref{app:analytic derivation of power spectrum}, with $\gamma=-4$ and $\log_{10} A = 0.5(\log_{10} Q_{\rm c} + 2.99)$ [see equation (\ref{eq:convert_Q_to_A})]. Hence the 90\% confidence interval $-29.6 \leq \log_{10} (Q_{\rm c} / 1 \, {\rm rad^2 \, s^{-3}} ) \leq -23.6$ in the second panel of Fig. \ref{fig:marginalised_posteriors_J1359-6038} converts into $-13.3 \leq \log_{10} A \leq -10.3$. This is marginally below the 95\% confidence interval measured independently by \citet{Lower2020} for PSR J1359$-$6038, viz.\ $\log_{10} A = -10.0^{+0.2}_{-0.1}$, noting a slight discrepancy at the 95\%-confidence level in the exponent, viz.\ $\gamma = -5.1^{+0.8}_{-0.9}$ versus $\gamma = -4$. More generally, the estimate in Fig. \ref{fig:marginalised_posteriors_J1359-6038} is comparable broadly with population studies of ordinary and millisecond pulsars, which yield $-15 \leq \log_{10} A \leq -4.9$ and $-20 \leq \gamma \leq -0.40$ \citep{2019MNRAS.489.3810P, Lower2020, 2023MNRAS.523.4603K}. It is also comparable to but higher than population studies of millisecond pulsars only, which yield $-17 \leq \log_{10} A \leq -12$ and $-7.5 \leq \gamma \leq -0.44$ \citep{2016MNRAS.458.2161L, 2020MNRAS.497.3264G, 2021MNRAS.502..478G}. The latter result is expected, as ordinary pulsars such as PSR J1359$-$6038 are known to be noisier typically than millisecond pulsars. 
Other statistics used in the literature to quantify timing noise strength such as $\sigma_z(\tau)$ \citep{1997A&A...326..924M, 2010MNRAS.402.1027H},  $\sigma_{\mathcal{R}}(m,T)$ \citep{1980ApJ...239..640C} and $\Delta_8(t)$ \citep{1994ApJ...422..671A, Shannon_2010}, are harder to compare to $Q_{\rm{c}}$ and are not compared here.

\begin{figure}
    \centering
    \includegraphics[width=0.45\textwidth]{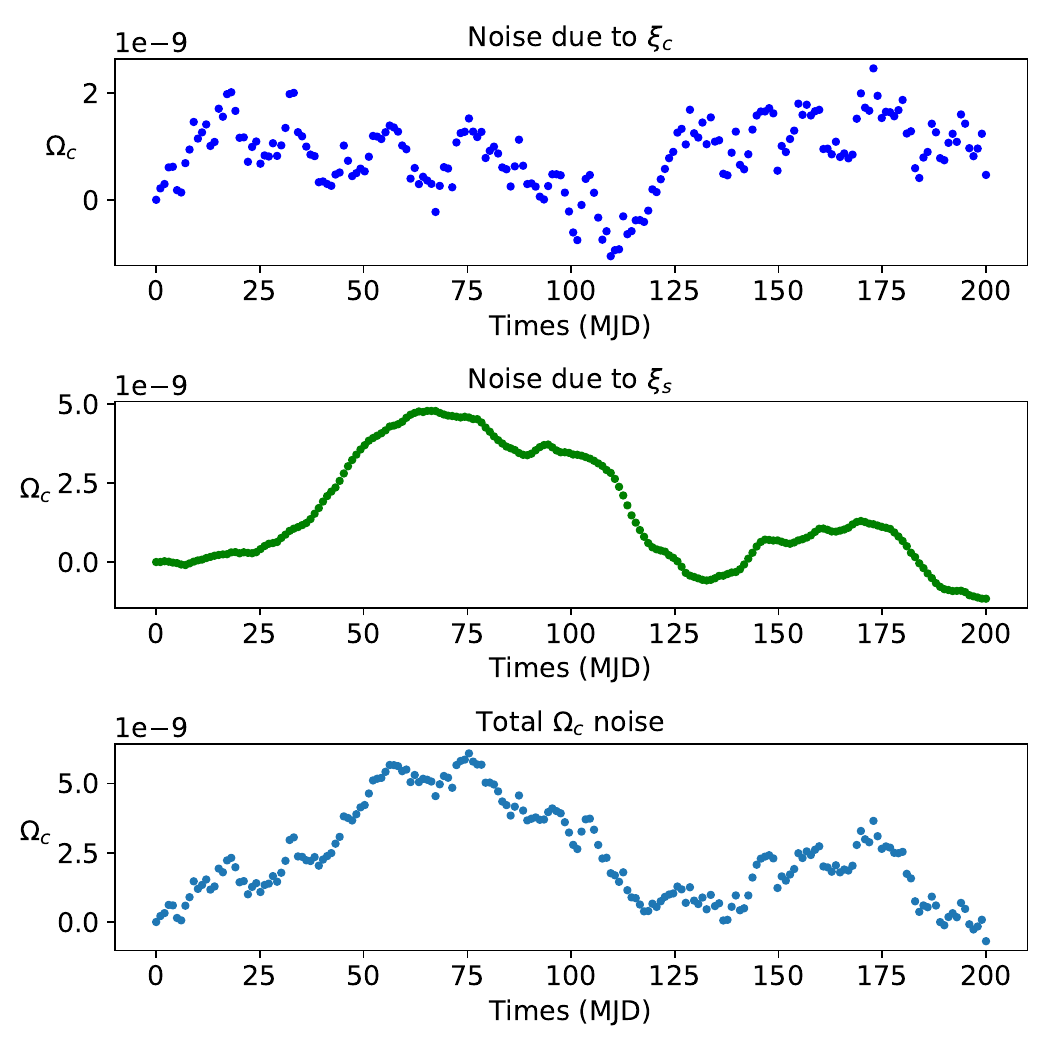}
    \caption{Residuals of simulated pulsar frequencies to illustrate the qualitative difference between the timing noise produced by $\xi_{\rm{c}}$ and $\xi_{\rm{s}}$. The first panel shows the fluctuations in $\Omega_{\rm{c}}$ for a simulated pulsar with $\xi_{\rm{c}} \neq 0$ but $\xi_{\rm{s}}=0$, the second panel shows $\Omega_{\rm{c}}$ produced for $\xi_{\rm{s}} \neq 0$ and $\xi_{\rm{c}} = 0$ and the third panel is for a simulated pulsar, where $\xi_{\rm{c}}$ and $\xi_{\rm{s}}$ are calculated by adding the fluctuations from the first and second panels.}
    \label{fig:compare_noise_types}
\end{figure}

The detailed shapes of the peaks in $Q_{\rm c}$ and $Q_{\rm s}$ and to a lesser extent their positions are influenced by the {\sc tempo2} fitting process which constructs local frequencies from TOAs, as described in Section \ref{sec:data}. To calibrate for this effect, we conduct Monte Carlo simulations with synthetic data comparing the $Q_{\rm c}$ and $Q_{\rm s}$ estimates inferred from local {\sc tempo2} TOA fits versus direct frequency data. The results are presented in Appendix \ref{app:tempo2 fits to local frequencies}.
When calculating local frequencies from TOAs, {\sc tempo2} fits a linear frequency model $\propto \dot \Omega_{\rm c} (t_n - t_{n-1})$, which does not make any allowance for a random walk in the interval $t_{n-1} \leq t \leq t_n$. Consequently {\sc tempo2} acts as a low-pass filter, which smooths out features on inter-TOA time-scales. If TOAs are spaced widely enough, so that one has $\tau \ll t_n - t_{n-1}$, then {\sc tempo2} smooths out some evidence for $\tau$ as well. 
Smoothing may also bias $Q_{\rm c}$ and $Q_{\rm s}$ \citep{Deeter_1984, Meyers2021a}. Because (\ref{eq:crust de}) and (\ref{eq:core de}) are linear differential equations, the timing noise in $\Omega_{\rm{c}}$ can be written as a linear combination of the timing noise from $\xi_{\rm{c}}$ and $\xi_{\rm{s}}$. These two contributions are separated and plotted in Fig. \ref{fig:compare_noise_types} for simulated data. The contribution from $\xi_{\rm{c}}$ is rougher than from $\xi_{\rm{s}}$. This is because $\xi_{\rm{c}}$ appears directly in equation (\ref{eq:crust de}) for $d \Omega_{\rm{c}}/dt$, whereas $\xi_{\rm{s}}$ only affects $\Omega_{\rm{c}}$ indirectly  through its time integral in $\Omega_{\rm{s}}$ in the coupling term.
Hence smoothing the data nullifies $\xi_{\rm c}$ fluctuations more than $\xi_{\rm s}$ fluctuations, overestimating $Q_{\rm{s}}$ and underestimating $Q_{\rm c}$.

\subsection{Average spin-down rate}
\label{ssec:average spin-down rate}

$\langle \dot \Omega_{\rm{c}} \rangle$ is clearly recoverable with a narrow peak. The rightmost panel of Fig. \ref{fig:marginalised_posteriors_J1359-6038} yields $-2.4489 \times 10^{-12} \leq \langle \dot{\Omega}_{\rm c} \rangle / (1 \, {\rm rad \,s^{-2}}) \leq -2.4465 \times 10^{-12}$ for the 90\% confidence interval, which is consistent with the value of $\dot{\Omega}_{\rm c}(t_1) = -2.4472 \times 10^{-12}~\rm{rad}~\rm{s}^{-2}$ predicted by \citet{Lower2020}. The narrow peak is expected; $\langle \dot \Omega_{\rm{c}} \rangle$ can be obtained by a linear regression in {\sc tempo2} without resorting to a Kalman filter, as demonstrated by decades of pulsar timing experiments. 

\subsection{Unidentifable parameters}
\label{ssec:unidentifiable parameters}

The marginalized posteriors for $r$ and $\langle \Omega_{\rm{c}} - \Omega_{\rm{s}} \rangle$ are not sharply peaked, as is apparent from columns 1 and 5 in the corner plot in Fig. \ref{fig:posteriors_J1359-6038} in Appendix \ref{app:full posterior distribution for PSR J1359-6038}. There is no evidence of railing against the prior bounds, but the marginalized posterior is flat and therefore uninformative; the probability at the extremes of the prior range is $\approx 80 \%$ and $\approx 90 \%$ of the peak probability for $r$ and $\langle \Omega_{\rm{c}} -\Omega_{\rm{s}} \rangle$ respectively.

The difficulty in recovering $r$ and $\langle \Omega_{\rm{c}} - \Omega_{\rm{s}} \rangle$ is predicted by the identifiability analysis in Section \ref{ssec:identifiability}. The deterministic part of equation (\ref{eq:em_only_omgddot}) does not feature $r$ and $\langle \Omega_{\rm c} - \Omega_{\rm s} \rangle$. Relatedly, the simulations in Appendix \ref{app:parameter estimation on simulated frequencies} show the recovery of the six model parameters on simulated $\Omega_{\rm c}$ data and confirm that $r$ and $\langle \Omega_{\rm{c}} - \Omega_{\rm{s}} \rangle$ are poorly recovered. The average distances of the recovered $\log_{10} r$ and $\langle \Omega_{\rm c} - \Omega_{\rm s} \rangle$ parameters in Fig. \ref{fig:simple_freq_test} from their true values as a fraction of their prior widths are $41\%$ and $32\%$ respectively.

\section{One-component model}
\label{sec:one-component model}

One may ask whether the two-component model described by Equations (\ref{eq:crust de}) and (\ref{eq:core de}) is needlessly elaborate, despite its sound phenomenological motivation. Can a simpler stochastic model explain the spin wandering statistics measured in PSR J1359$-$6038? The question becomes especially pertinent, when one acknowledges the difficulty in identifying $r$, $Q_{\rm{s}}$ and $\langle \Omega_{\rm{c}}-\Omega_{\rm{s}} \rangle$, as seen in Figs. \ref{fig:simple_freq_test} and \ref{fig:posteriors_J1359-6038} in Appendices \ref{ssec:simulation under ideal conditions} and \ref{app:full posterior distribution for PSR J1359-6038}. It is possible, at least in principle, that the challenges of identification arise because the model is more complicated than it needs to be. 

To test the above hypothesis, we repeat the Kalman filter analysis in Sections \ref{sec:kalman tracking and estimation}--\ref{sec:results} for a one-component model. The single component corresponds to the crust (subscript `c'), which is phase-locked to the radio pulsations. The one-component equation of motion takes the form $I_{\rm c} d\Omega_{\rm{c}}/dt = N_{\rm{c}} + \xi_{\rm{c}}(t)$, analogous to Equation (\ref{eq:crust de}) but without the crust-superfluid coupling, where $\xi_{\rm{c}}(t)$ is a Langevin driving torque; see Appendix \ref{app:one-component model} for details. Parameter estimates for the model parameters, namely $Q_{\rm{c}}$ and $\langle \dot \Omega_{\rm{c}} \rangle$, are presented in the format of a traditional corner plot in Appendix \ref{app:one-component model}; see Fig. \ref{fig:posteriors_J1359-6038_onecomp} and Table \ref{tab:onecomp_result_table}. The posterior peaks at $\log_{10} Q_{\rm{c}}/(1~\rm{rad}^2\rm{s}^{-3}) = -22.73^{+0.23}_{-0.19}$  and $\langle \dot \Omega_{\rm{c}} \rangle = -2.4473 \times 10^{-12} \pm 7 \times 10^{-16} ~\rm{rad}~\rm{s}^{-2}$ (90\% confidence intervals). 

The $\langle \dot \Omega_{\rm{c}} \rangle$ value recovered above is similar to that recovered for the two-component model, because it is insensitive to how the timing noise is modelled. 
However, the recovered $Q_{\rm{c}}$ for the one-component model is larger than for the two-component model by a factor of $10^{1.3} \approx 20$. This is likely because $Q_{\rm c}$ and $Q_{\rm s}$ combine through the crust-superfluid coupling to generate the noise in $\Omega_{\rm c}(t)$ in the two-component model, whereas only $Q_{\rm c}$ is responsible in the one-component model, and we find $Q_{\rm s} \sim 10^2 Q_{\rm c}$ in Fig. \ref{fig:marginalised_posteriors_J1359-6038} and Table \ref{tab:twocomp_result_table} for the two-component model.

We can compare $Q_{\rm{c}}$ for the one-component model to the PSD normalisation $A$ measured by other researchers as in Section \ref{ssec:torque_noise_amplitudes}. The one-component estimate of $Q_{\rm{c}}$ in Table \ref{tab:onecomp_result_table} converts to $\log_{10} A = -9.87^{+0.12}_{-0.09}$. This is consistent with the estimate for PSR J1359$-$6038 in \citet{Lower2020} and with estimates for ordinary and millisecond pulsars more broadly \citep{2019MNRAS.489.3810P, Lower2020, 2023MNRAS.523.4603K}.

A Bayesian model comparison shows that the two-component model is strongly preferred in a Bayesian sense, with a $\log_{10}$ Bayes factor of $6.81 \pm 0.02$ relative to the one-component model. Further details can be found in Appendix \ref{app:one-component model}.

\section{Conclusion}
\label{sec:conclusion}

Traditionally, timing noise studies proceed by comparing the measured variance in the phase residuals to that predicted by a microphysical or phenomenological model \citep{Boynton_1972, Groth_1975_III, Cordes_1980, 1980ApJ...239..640C, Shannon_2010, Melatos2014}, fitting a microphysical or phenomenological model to the power spectral density of the phase residuals \citep{Van_Haasteren_2009, 2010MNRAS.402.1027H, 2016MNRAS.458.2161L, 2019MNRAS.489.3810P, 2020MNRAS.494.2012P, 2020MNRAS.497.3264G}, or measuring a relaxation time-scale using the autocorrelation function of the phase residuals \citep{Price_2012}. 
These approaches raise interesting questions about the type of variance measured, e.g. Allan variance \citep{1997A&A...326..924M, 2010MNRAS.402.1027H, Shannon_2010, Melatos2014}, or biases in constructing the power spectral density \citep{2011MNRAS.418..561C, 2013MNRAS.428.1147V, 2023MNRAS.523.4603K}.
In this paper we fit a model directly to the time-ordered $\Omega_{\rm c}(t)$ data without averaging implicitly over an ensemble, i.e. without analyzing the phase residual PSD. The extra information made available by the analysis of a unique, time-ordered, random realization makes it feasible to estimate reliably four of the six static parameters in the classic, two-component, crust-superfluid model of a neutron star, and to distinguish statistically between one- and two-component models, with interesting astrophysical implications.

In this paper, the Kalman filter parameter estimation method developed originally by \citet{Meyers2021a, Meyers2021b} is applied to real astronomical data from a single, accurately timed object, namely PSR J1359$-$6038, with the aim of demonstrating the practical effectiveness of the method. 
The posterior distribution for the six two-component model parameters is shown in Fig. \ref{fig:posteriors_J1359-6038}. Four of the six parameters are recovered reliably from the data. The results are summarised in Fig. \ref{fig:marginalised_posteriors_J1359-6038} and Table \ref{tab:twocomp_result_table}. The peak estimates are $\tau = 1.3 \times 10^7~\rm{s}$, $Q_{\rm{c}} = 1.05 \times 10^{-24}~\rm{rad^2~s^{-3}}$, $Q_{\rm{s}} = 1.82 \times 10^{-22}~\rm{rad^2~s^{-3}}$ and $\langle \dot \Omega_{\rm{c}} \rangle$ = $-2.4475 \times 10^{-12} ~\rm{rad~s^{-2}}$. The associated $90\%$ confidence intervals are $6.6 \leq \log_{10} \tau/(1~\rm{s}) \leq 7.9$, $-29.6 \leq \log_{10} Q_{\rm{c}}/(1~\rm{rad^2~s^{-3}}) \leq -23.6$, $-22.6 \leq \log_{10}Q_{\rm{s}}/(1~\rm{rad^2~s^{-3}}) \leq -18.2$ and $-2.4489 \times 10^{-12}~{\rm{rad~s^{-2}}} \leq \langle \dot \Omega_{\rm{c}} \rangle \leq -2.4465 \times 10^{-12}~\rm{rad~s^{-2}}$. 
The inferred coupling time-scale is broadly consistent with independent measurements based on auto-correlating phase residuals \citep{Price_2012} or fitting exponential post-glitch recoveries \citep{2022MNRAS.511..425G}. The torque noise amplitudes are broadly consistent with independent measurements based on the phase residuals PSD \citep{2019MNRAS.489.3810P, Lower2020, 2023MNRAS.523.4603K}. Two of the six two-component model parameters, namely $r$ and $\langle \Omega_{\rm{c}} - \Omega_{\rm{s}} \rangle$, cannot be measured reliably, in line with the formal identifiability analysis in Section \ref{ssec:identifiability} and Appendix \ref{app:identifiability of Qc, Qs and R}.
The estimates of $\tau$, $Q_{\rm c}$, and $Q_{\rm s}$, once extended to more pulsars, are likely to help illuminate the physical origin of timing noise and the nature of the spin-down-driven, far-from-equilibrium mechanisms which drive stochasticity in neutron star interiors, such as starquakes \citep{2006ApJ...652.1531M, 2010MNRAS.407L..54C, 2022MNRAS.511.3365G, 2022MNRAS.514.1628K}, superfluid instabilities \citep{2003PhRvL..90i1101A, Melatos2014}, magnetospheric fluctuations \citep{1987ApJ...321..799C}, and superfluid vortex avalanches \citep{1975Natur.256...25A, 2011MNRAS.415.1611W}. 

Bayesian model selection shows that the classic two-component model in Section \ref{sec:two-component model} fits the data better than the representative one-component model given by equations (\ref{eq:onecomp_torque})-(\ref{eq:onecomp_noise}), with a $\log_{10}$ Bayes factor of $6.81 \pm 0.02$. 
The Kalman filter’s ability to recover successfully $\tau$, $Q_{\rm c}$, and $Q_{\rm s}$ for the two-component model, noting that $\tau$ and $Q_{\rm s}$ do not feature in the one-component model, adds support to the conclusion of the model selection exercise. 

The results in this paper exemplify the usefulness in astrophysics of parameter estimation methods based on Kalman filtering and similar algorithms \citep{Meyers2021a}. In the future, more data and more sophisticated models will give more accurate parameter estimates.
The Kalman filter can be rewritten easily to track using physical models other than equations (\ref{eq:crust de}) and (\ref{eq:core de}). A nonlinear torque can be incorporated to calculate the braking index, if the data volume is sufficient \citep{2023MNRAS.522.4880V}. The model currently assumes that $\xi_{\rm{c}}$ and $\xi_{\rm{s}}$ are white noise torques, which implies that the PSD of $\Omega_{\rm c}$ fluctuations scales as $f^{-4}$ at large $f$, a property which is satisfied approximately by some but not all pulsars \citep{Lower2020, 2023MNRAS.520.2813A, 2023MNRAS.523.4603K}. Generalizing $\xi_{\rm c}$ and $\xi_{\rm s}$ to colored noise is a standard procedure in electrical engineering \citep{Gelb1974}. Bayesian model selection between these physically motivated alternatives is straightforward too, because the Kalman tracker and nested sampler in this paper together generate the Bayesian evidence as a by-product of the analysis.

The next step in this investigation is to extend the Kalman tracker so that it operates on TOAs directly instead of converting them first to local frequencies $\Omega_{\rm c}(t_i)$ using {\sc tempo2}. Monte Carlo simulations in Appendix \ref{app:tempo2 fits to local frequencies} show that local {\sc tempo2} computation of $\Omega_{\rm c}(t_i)$ biases the static parameter estimation results, underestimating $Q_{\rm c}$ and making $\tau$ harder to infer. This is not a fault with {\sc tempo2}; it is a general consequence of the low-pass filtering introduced by any local frequency fitting process. Once the Kalman tracker is extended to operate on TOAs directly, it will be appropriate to apply the method to a wider selection of pulsars beyond PSR J1359$-$6038 and conduct population studies of the recoverable quantities $\tau$, $Q_{\rm c}$, and $Q_{\rm s}$, which are important physically and are measured in only a few objects to date. 
We note in closing that the Kalman tracker and nested sampler in this paper are easy to implement and quick to run. By way of calibration, the PSR J1359$-$6038 analysis in this paper takes of order one hour to run on $\sim 10^3$ TOAs. More generally, the run time scales in direct proportion to the number of TOAs.

\section*{Acknowledgements}

The authors thank Liam Dunn for discussions on using the {\sc tempo2} software to analyse pulsar timing data. Parts of this research were conducted by the Australian Research Council Centre of Excellence for Gravitational Wave Discovery (OzGrav), through project number CE170100004. NJO is the recipient of a Melbourne Research Scholarship. The numerical calculations were performed on the OzSTAR supercomputer facility at Swinburne University of Technology. The OzSTAR program receives funding in part from the Astronomy National Collaborative Research Infrastructure Strategy (NCRIS) allocation provided by the Australian Government.

\section*{Data Availability}

The pulsar timing data for PSR J1359$-$6038 come from \citet{Lower2020} and are available at \url{https://github.com/Molonglo/TimingDataRelease1/}. We use the {\sc tempo2} software package \citep{Edwards_2006a, Edwards_2006b} to process the real and synthetic data. The software for applying the Kalman filter-based parameter estimation method discussed in this paper to real pulsar data and for carrying out simulations with this method is available at \url{https://github.com/oneill-academic/pulsar_freq_filter}.


\bibliographystyle{mnras}
\bibliography{references_ads,references_non_ads}



\appendix

\section{Power spectrum}
\label{app:analytic derivation of power spectrum}

\noindent
In this appendix we derive the power spectral density of the dependent variables $\Omega_{\rm c}$ and $\Omega_{\rm s}$ in the two-component model given by the differential equations (\ref{eq:crust de}) and (\ref{eq:core de}). 
The behaviour and power spectrum of the stochastic part of the solutions are not affected by the constant torques so they are removed from the differential equations for this calculation. This is equivalent to subtracting a linear best fit and analysing the residuals. A similar calculation appears in Appendix A of \citet{Meyers2021b} and has been generalised to more than two stellar components by \citet{2023MNRAS.520.2813A}.

Upon Fourier transforming (\ref{eq:crust de}) and (\ref{eq:core de}), one obtains
\begin{align}
    i \omega \hat{\Omega}_{\rm{c}}(\omega) &= - \frac{1}{\tau_{\rm c}} \hat{\Omega}_{\rm{c}}(\omega) + \frac{1}{\tau_{\rm c}} \hat{\Omega}_{\rm{s}}(\omega) + \frac{\hat{\xi}_{\rm{c}}(\omega)}{I_{\rm c}},\label{eq:c_fourier_transformed}\\
    i \omega \hat{\Omega}_{\rm{s}}(\omega) &= - \frac{1}{\tau_{\rm s}} \hat{\Omega}_{\rm{s}}(\omega) + \frac{1}{\tau_{\rm s}} \hat{\Omega}_{\rm{c}}(\omega) + \frac{\hat{\xi}_{\rm{s}}(\omega)}{I_{\rm s}} \label{eq:s_fourier_transformed}.
\end{align}

\noindent
Solving (\ref{eq:c_fourier_transformed}) and (\ref{eq:s_fourier_transformed}) for $ \hat{\Omega}_{\rm{c}}(\omega)$ and $ \hat{\Omega}_{\rm{s}}(\omega)$ gives

\begin{align}
\hat{\Omega}_{\rm{c}}(\omega) &= \frac{\left( i \omega +\frac{1}{\tau_{\rm s}}\right) \frac{\hat{\xi}_{\rm{c}}(\omega)}{I_{\rm c}} + \frac{1}{\tau_{\rm c}}\frac{\hat{\xi}_{\rm{s}}(\omega)}{I_{\rm s}}}{-\omega^2 + i \omega/\tau}, \label{eq:omegac_fourier_transform}\\
\hat{\Omega}_{\rm{s}}(\omega) &= \frac{\frac{1}{\tau_{\rm s}}\frac{\hat{\xi}_{\rm{c}}(\omega)}{I_{\rm c}} + \left( i \omega +\frac{1}{\tau_{\rm c}}\right) \frac{\hat{\xi}_{\rm{s}}(\omega)}{I_{\rm s}}}{-\omega^2 + i \omega/\tau}. \label{eq:omegas_fourier_transform}
\end{align}

Given two random variables $x(t)$ and $y(t)$, their cross power spectral density $P_{xy}(\omega)$ (or simply the power spectral density if $x=y$) can be calculated by the formula
\begin{align}
\langle \hat x^{*}(\omega) \hat y(\omega') \rangle &= 2\pi \delta(\omega-\omega')P_{xy}(\omega).
\end{align}

\noindent
In this paper it is assumed that $\xi_{\rm{c}}/I_{\rm c}$ and $\xi_{\rm{s}}/I_{\rm s}$ are uncorrelated and stationary white noise processes with flat spectra, so we have
\begin{align}
P_{\xi_{\rm{c}}\xi_{\rm{c}}}(\omega) &= \sigma_{\rm{c}}^2,\label{eq:xi_c_spectrum}\\
P_{\xi_{\rm{s}}\xi_{\rm{s}}}(\omega) &= \sigma_{\rm{s}}^2,\label{eq:xi_s_spectrum}\\
P_{\xi_{\rm{c}}\xi_{\rm{s}}}(\omega) &= 0.\label{eq:xi_c_xi_s_spectrum}
\end{align}

\noindent
Combining (\ref{eq:xi_c_spectrum})-(\ref{eq:xi_c_xi_s_spectrum}) with (\ref{eq:omegac_fourier_transform}) and (\ref{eq:omegas_fourier_transform}) gives

\begin{align}
\label{eq:app:omgc_psd_full}
P_{\Omega_{\rm{c}}\Omega_{\rm{c}}}(\omega) &= \frac{\left(\omega^2 +\frac{1}{\tau_{\rm s}^2}\right) \frac{\sigma_{\rm{c}}^2}{I_{\rm c}^2} + \frac{1}{\tau_{\rm c}^2}\frac{\sigma_{\rm{s}}^2}{I_{\rm s}^2}}{\omega^4 + \omega^2/\tau^2},\\
\label{eq:app:omgs_psd_full}
P_{\Omega_{\rm{s}}\Omega_{\rm{s}}}(\omega) &= \frac{\left(\omega^2 +\frac{1}{\tau_{\rm c}^2}\right) \frac{\sigma_{\rm{s}}^2}{I_{\rm s}^2} + \frac{1}{\tau_{\rm s}^2}\frac{\sigma_{\rm{c}}^2}{I_{\rm c}^2}}{\omega^4 + \omega^2/\tau^2},\\
\label{eq:app:omgcs_psd_full}
P_{\Omega_{\rm{c}}\Omega_{\rm{s}}}(\omega) &= \frac{i \omega \left( \frac{\sigma_{\rm{s}}^2}{\tau_{\rm c} I_{\rm s}^2} - \frac{\sigma_{\rm{c}}^2}{\tau_{\rm s} I_{\rm c}^2} \right) + \left( \frac{\sigma_{\rm{c}}^2}{\tau_{\rm s}^2 I_{\rm c}^2}+  \frac{\sigma_{\rm{s}}^2}{\tau_{\rm c}^2 I_{\rm s}^2} \right)}{\omega^4 + \omega^2/\tau^2}.
\end{align}
Equations (\ref{eq:app:omgc_psd_full}) and (\ref{eq:app:omgs_psd_full}) asymptote towards pure power laws as functions of $\omega$ in the limits of high and low $\omega$. Specifically, the PSDs of $\Omega_{\rm c}$ and $\Omega_{\rm s}$ tend to a spectral index of $-2$ at high and low $\omega$; see \citet{Meyers2021b} for more details.

In the literature, timing noise is often described by the power spectrum of timing residuals, which is often written in the form
\begin{align}
P(f) = \frac{A^2}{12\pi^2 f_{\rm{yr}}^{3}} \left(\frac{f}{f_{\rm{yr}}}\right)^{\gamma} \label{eq:gw_spectrum2}
\end{align}
\citep{2016MNRAS.458.2161L}. 
In the limit $\omega \gg 1/\tau$, equation (\ref{eq:app:omgc_psd_full}) implies 
\begin{align}
P(f) &= \frac{Q_{\rm c}}{(2\pi)^4 \Omega_0^2 f^4}, \label{eq:spectrum_limit}
\end{align}
noting that $P(f) = P_{\Omega_{\rm{c}}\Omega_{\rm{c}}}(\omega)/(\Omega_0^2 \omega^2)$, where $\Omega_0 = 49.28 ~\rm{rad}~\rm{s}^{-1}$ is the pulsar's angular frequency. If we set $\gamma = -4$ in equation (\ref{eq:gw_spectrum2}) we can convert $Q_{\rm c}$ to $A^2$ by the formula
\begin{align}
A^2 &= \frac{12\pi^2}{(2\pi)^4 \Omega_0^{2} f_{\rm{yr}}} Q_{\rm c}.
\end{align}
Numerically we have
\begin{align}
\log_{10} A &= 0.5(\log_{10} Q_{\rm c} + 2.99). \label{eq:convert_Q_to_A}
\end{align}
In equation (\ref{eq:convert_Q_to_A}) and elsewhere, $A$ is unitless and $Q_{\rm c}$ has units of $\rm{rad}^2\rm{s}^{-3}$.

\section{State Space Representation}
\label{app:full state space representation}

In this appendix we give the full forms of the Kalman filter update matrices $\boldsymbol F_i$, $\boldsymbol T_i$, and $\boldsymbol Q_i$ defined in equations (\ref{eq:kalman update}) and (\ref{eq:process noise cov}) to assist the interested reader in reproducing the numerical results in Section \ref{sec:results}. Specifically, we have
\begin{align}
\boldsymbol F_i 
&=  
\frac{1}{\taus + \tauc}
\begin{pmatrix} 
\tauc+\taus e^{-\Delta t_i/\tau} & \taus - \taus e^{- \Delta t_i/\tau}\\ 
\tauc-\tauc e^{-\Delta t_i/\tau} & \taus + \tau_{\rm c} e^{-\Delta t_i/\tau}
\end{pmatrix},
\label{eq:transition_matrix_full}
\end{align}

\begin{align}
\boldsymbol N_i 
&= 
\langle \dot \Omega_{\rm c} \rangle  \Delta t_i
\begin{pmatrix} 
1\\
1
\end{pmatrix}
+
\langle \Omega_{\rm c} - \Omega_{\rm s} \rangle \left(1 - e^{-\Delta t_i/\tau}\right)
\begin{pmatrix} 
\tau/\tauc \\
-\tau/\taus
\end{pmatrix},
\label{eq:torque_vector_full}
\end{align}

\begin{align}
\boldsymbol Q_i &= \left(\frac{1}{\tau_{\rm c} + \tau_{\rm s}}\right)^2
\begin{pmatrix}
a & b\\
b & c
\end{pmatrix},
\label{eq:process_noise_covariance_full}
\end{align}
with
\begin{align}
\nonumber a=&\left(\frac{\sigma_{\rm{c}}^2\tauc^2}{I_{\rm{c}}^2} + \frac{\sigma_{\rm{s}}^2\taus^2}{I_{\rm{s}}^2}\right)\Delta t_i\\
\nonumber&+ 2\tau \taus \left(\frac{\sigma_{\rm{c}}^2\tauc}{I_{\rm{c}}^2} - \frac{\sigma_{\rm{s}}^2\taus}{I_{\rm{s}}^2}\right)\left(1 - e^{-\Delta t_i/\tau}\right)\\
 &+\frac{\tau\taus^2}{2}\left(\frac{\sigma_{\rm{c}}^2}{I_{\rm{c}}^2} + \frac{\sigma_{\rm{s}}^2}{I_{\rm{s}}^2}\right)\left(1 - e^{-2\Delta t_i/\tau}\right),\\
\nonumber b=&\left(\frac{\sigma_{\rm{c}}^2\tauc^2}{I_{\rm{c}}^2} + \frac{\sigma_{\rm{s}}^2\taus^2}{I_{\rm{s}}^2}\right)\Delta t_i\\
\nonumber&+ \tau(\tau_{\rm s}-\tau_{\rm c})\left( \frac{\sigma_{\rm{c}}^2\tau_{\rm c}}{I_{\rm{c}}^2} - \frac{\sigma_{\rm{s}}^2\tau_{\rm s}}{I_{\rm{s}}^2} \right)\left(1 - e^{-\Delta t_i/\tau}\right)\\
&-\frac{\tau\tauc\taus}{2}\left(\frac{\sigma_{\rm{c}}^2}{I_{\rm{c}}^2} + \frac{\sigma_{\rm{s}}^2}{I_{\rm{s}}^2}\right)\left(1 - e^{-2\Delta t_i/\tau}\right),\\
\nonumber c=&\left(\frac{\sigma_{\rm{c}}^2\tauc^2}{I_{\rm{c}}^2} + \frac{\sigma_{\rm{s}}^2\taus^2}{I_{\rm{s}}^2}\right)\Delta t_i\\
\nonumber&+ 2\tau\tauc\left(\frac{\sigma_{\rm{s}}^2\taus}{I_{\rm{s}}^2} - \frac{\sigma_{\rm{c}}^2\tauc}{I_{\rm{c}}^2}\right)\left(1 - e^{-\Delta t_i/\tau}\right)\\
&+\frac{\tau\tauc^2}{2}\left(\frac{\sigma_{\rm{c}}^2}{I_{\rm{c}}^2} + \frac{\sigma_{\rm{s}}^2}{I_{\rm{s}}^2}\right)\left(1 - e^{-2\Delta t_i/\tau}\right).
\end{align}

\section{Kalman filter}
\label{app:explanation of the Kalman filter}

In this appendix, for the sake of completeness and reproducibility, we summarize the structure and operation of the Kalman filter used to generate the results in Sections \ref{sec:results} and \ref{sec:one-component model}. The discussion includes a short justification of the form of the Kalman likelihood in equation (\ref{eq:kalman_likelihood}).

The Kalman filter is a predictor-corrector algorithm. Given a sequence of $i-1$ measurements, $\boldsymbol Y_1, \boldsymbol Y_2, ...,\boldsymbol Y_{i-1}$, the Kalman filter predicts the value of the state $\boldsymbol X_i$ at the next time step $t_i$. The estimate is denoted by $\boldsymbol X_i^{i-1}$, and its error is given by the covariance matrix $\boldsymbol P_i^{i-1}$. When the measurement $\boldsymbol Y_i$ is made, the estimate $\boldsymbol X_i^{i-1}$ is updated to give the new estimate of the current state, $\boldsymbol X_i^i$, and its error, $\boldsymbol P_i^i$. This procedure is carried out from $t_1$ to $t_N$. 

The following update formulas (\ref{kalman_initial_state})--(\ref{kalman_cov_update}) implement the prediction and correction steps described above. The symbols are explained in the text immediately following. 

Initialisation:
\begin{align}
\label{kalman_initial_state}
\boldsymbol X_0^0 &= \boldsymbol \mu_0, \\
\label{kalman_initial_cov}
\boldsymbol P_0^0 &= \boldsymbol S_0.
\end{align}

State prediction:
\begin{align}
\label{kalman_state_estimate}
\boldsymbol X_i^{i-1} &= \boldsymbol F_i \boldsymbol X_{i-1}^{i-1} + \boldsymbol T_i,\\
\label{kalman_cov_estimate}
\boldsymbol P_i^{i-1} &= \boldsymbol F_i \boldsymbol P^{i-1}_{i-1} \boldsymbol F^T_i + \boldsymbol Q_i.
\end{align}

State correction:
\begin{align}
\label{kalman_error}
\boldsymbol \epsilon_i &= \boldsymbol Y_i - E(\boldsymbol Y_i \vert \boldsymbol Y_{1:i-1}) = \boldsymbol Y_i - \boldsymbol C \boldsymbol X_i^{i-1}, \\
\label{kalman_error_cov}
\boldsymbol S_i &= \boldsymbol C \boldsymbol P_i^{i-1} \boldsymbol C^T + \boldsymbol R_i, \\
\label{kalman_gain}
\boldsymbol K_i &= \boldsymbol P_i^{i-1} \boldsymbol C^T \boldsymbol S_i^{-1}, \\
\label{kalman_state_update}
\boldsymbol X_i^i &= \boldsymbol X_i^{i-1} + \boldsymbol K_i \boldsymbol \epsilon_i, \\
\label{kalman_cov_update}
\boldsymbol P_i^i &= (1 - \boldsymbol K_i \boldsymbol C) \boldsymbol P_i^{i-1}.
\end{align}

\noindent
Equations (\ref{kalman_initial_state}) and (\ref{kalman_initial_cov}) estimate the initial state and its error. Equations (\ref{kalman_state_estimate}) and (\ref{kalman_cov_estimate}) predict the next state from the previous data. Equations (\ref{kalman_error}) through (\ref{kalman_cov_update}) combine the new measurement with the prediction to get a new estimate of the current state. In (\ref{kalman_error}), we define $\boldsymbol \epsilon_i = \boldsymbol Y_i - E(\boldsymbol Y_i \vert \boldsymbol Y_{1:i-1})$. In (\ref{kalman_error_cov}), we define $\boldsymbol S_i=\textrm{var}(\boldsymbol \epsilon_i)$. The matrix $\boldsymbol K_i$ in (\ref{kalman_gain})--(\ref{kalman_cov_update}) is usually called the Kalman gain.

To get the likelihood, $p(\boldsymbol Y = \boldsymbol Y_{1:N} | \boldsymbol \theta)$, from the Kalman filter estimates, we apply standard rules for conditional probability to get

\begin{align}
p(\boldsymbol Y_{1:N} \vert \boldsymbol \theta) &= p(\boldsymbol Y_N \vert \boldsymbol Y_{1:N-1}, \boldsymbol \theta) p(\boldsymbol Y_{1:N-1} | \boldsymbol \theta)\\
&= p(\boldsymbol Y_N \vert \boldsymbol Y_{1:N-1}, \boldsymbol \theta) p(\boldsymbol Y_{N-1} | \boldsymbol Y_{1:N-2}, \boldsymbol \theta)p(\boldsymbol Y_{1:N-2} | \boldsymbol \theta)
\end{align}
and hence recursively
\begin{align}
p(\boldsymbol Y_{1:N} \vert \boldsymbol \theta) &= \prod_{i=1}^N p(\boldsymbol Y_i | \boldsymbol Y_{1:i-1}, \boldsymbol \theta). \label{eq:product of pdfs}
\end{align}

\noindent
$p(\boldsymbol Y_i \vert \boldsymbol Y_{1:i-1}, \boldsymbol \theta)$ is Gaussian if all the errors are assumed to be Gaussian. The mean and covariance matrix of the $p(\boldsymbol Y_{1:N} \vert \boldsymbol \theta)$ distribution are
\begin{align}
E(\boldsymbol Y_i \vert \boldsymbol Y_{1:i-1}) &= \boldsymbol C \boldsymbol X_i^{i-1},\\
\textrm{var}(\boldsymbol Y_i \vert \boldsymbol Y_{1:i-1}) &= \boldsymbol S_i,
\end{align}

\noindent
which imply
\begin{align}
p(\boldsymbol Y_i \vert \boldsymbol Y_{1:i-1}, \boldsymbol \theta) = \mathcal{N}(\boldsymbol Y_i; \boldsymbol C \boldsymbol X_i^{i-1}, \boldsymbol S_i),
\end{align}
where $\mathcal{N}(\boldsymbol X; \boldsymbol \mu, \boldsymbol \Sigma)$ denotes a Gaussian with mean $\boldsymbol \mu$ and covariance matrix $\boldsymbol \Sigma$. Equivalently, writing $\boldsymbol \epsilon_i = \boldsymbol Y_i- \boldsymbol C \boldsymbol X_i^{i-1}$, we obtain
\begin{align}
p(\boldsymbol Y_i \vert \boldsymbol Y_{1:i-1}, \boldsymbol \theta) &= p(\boldsymbol \epsilon_i \vert \boldsymbol Y_{1:i-1}, \boldsymbol \theta)\\
&=\mathcal{N}(\boldsymbol \epsilon_i; 0, \boldsymbol S_i).\label{eq:pdf of error}
\end{align}

\noindent
The full formula for the log-likelihood then becomes
\begin{align}
\log p(\boldsymbol Y_{1:N} \vert \boldsymbol \theta) &= \sum_{i=1}^N \log p(\boldsymbol Y_i \vert \boldsymbol Y_{1:i-1}, \boldsymbol \theta) \label{eq:sum of lls}\\
&= -\frac{1}{2} \sum_{i=1}^N \left[ N_{\boldsymbol Y} \log(2\pi) + \log \det (\boldsymbol{S}_{i}) + \boldsymbol{\epsilon}_{i}^T \boldsymbol{S}_{i}^{-1} \boldsymbol{\epsilon}_{i} \right],\label{eq:kalman_likelihood2}
\end{align}
where $N_{\boldsymbol Y}$ is the dimension of $\boldsymbol Y$. Equation (\ref{eq:sum of lls}) follows from (\ref{eq:product of pdfs}) and (\ref{eq:kalman_likelihood2}) follows from (\ref{eq:pdf of error}). Equation (\ref{eq:kalman_likelihood2}) is the same as equation (\ref{eq:kalman_likelihood}) in Section \ref{ssec:kalman filter likelihood}.

Once $p(\boldsymbol Y_{1:N} | \boldsymbol \theta)$ is known, Bayes's theorem can be used to get the probability of the parameters in terms of the data,
\begin{align}
p(\boldsymbol \theta | \boldsymbol Y_{1:N}) &= \frac{p(\boldsymbol Y_{1:N} | \boldsymbol \theta)p(\boldsymbol \theta)}{p(\boldsymbol Y_{1:N})}. \label{bayes_theorem}
\end{align}

\section{Parameter estimation with simulated frequencies}
\label{app:parameter estimation on simulated frequencies}

\subsection{Simulation under ideal conditions}
\label{ssec:simulation under ideal conditions}

In this appendix our parameter estimation method is tested on simulated frequency data. 
To simulate the data, we must first select values for the model parameters $\tau_{\rm{c}}, \tau_{\rm{s}}, \sigma_{\rm{c}}/I_{\rm{c}}, \sigma_{\rm{s}}/I_{\rm{s}}, N_{\rm{c}}/I_{\rm{c}}, N_{\rm{s}}/I_{\rm{s}}$. We then chose an initial value $\Omega_{\rm{c}}(0)$ and set $\Omega_{\rm{s}}(0)$ to be
\begin{align}
\Omega_{\rm{s}}(0) &= \Omega_{\rm{c}}(0) - \langle \Omega_{\rm{c}} - \Omega_{\rm{s}} \rangle\\
&= \Omega_{\rm{c}}(0) - \tau \left( \frac{N_{\rm{c}}}{I_{\rm{c}}} - \frac{N_{\rm{s}}}{I_{\rm{s}}} \right)
\end{align}
so the pulsar is initially in equilibrium.
We choose $N_{\rm obs}$ random times from $t=0$ to $t=T_{\rm obs}$, where simulated measurements occur. The differential equations (\ref{eq:crust de}) and (\ref{eq:core de}) are then integrated from the initial time to each measurement time, giving $\Omega_{\rm{c}}(t_i)$ and $\Omega_{\rm{s}}(t_i)$ values at each $1 \leq i \leq N_{\rm obs}$. Realistic timing experiments only yield $\Omega_{\rm{c}}$ observations, so the $\Omega_{\rm{s}}$ values are discarded. 
To simulate measurement errors we add a number drawn from a Gaussian with variance $R$ to each $\Omega_{\rm{c}}$ value. For simplicity in these simulations we assume every data point has the same measurement uncertainty, unlike in real data.
Once the data are simulated, the parameter estimation algorithm is executed, and the posterior distribution is compared to the true parameters. 

\begin{figure*}
    \centering
\includegraphics[width=0.95\textwidth]{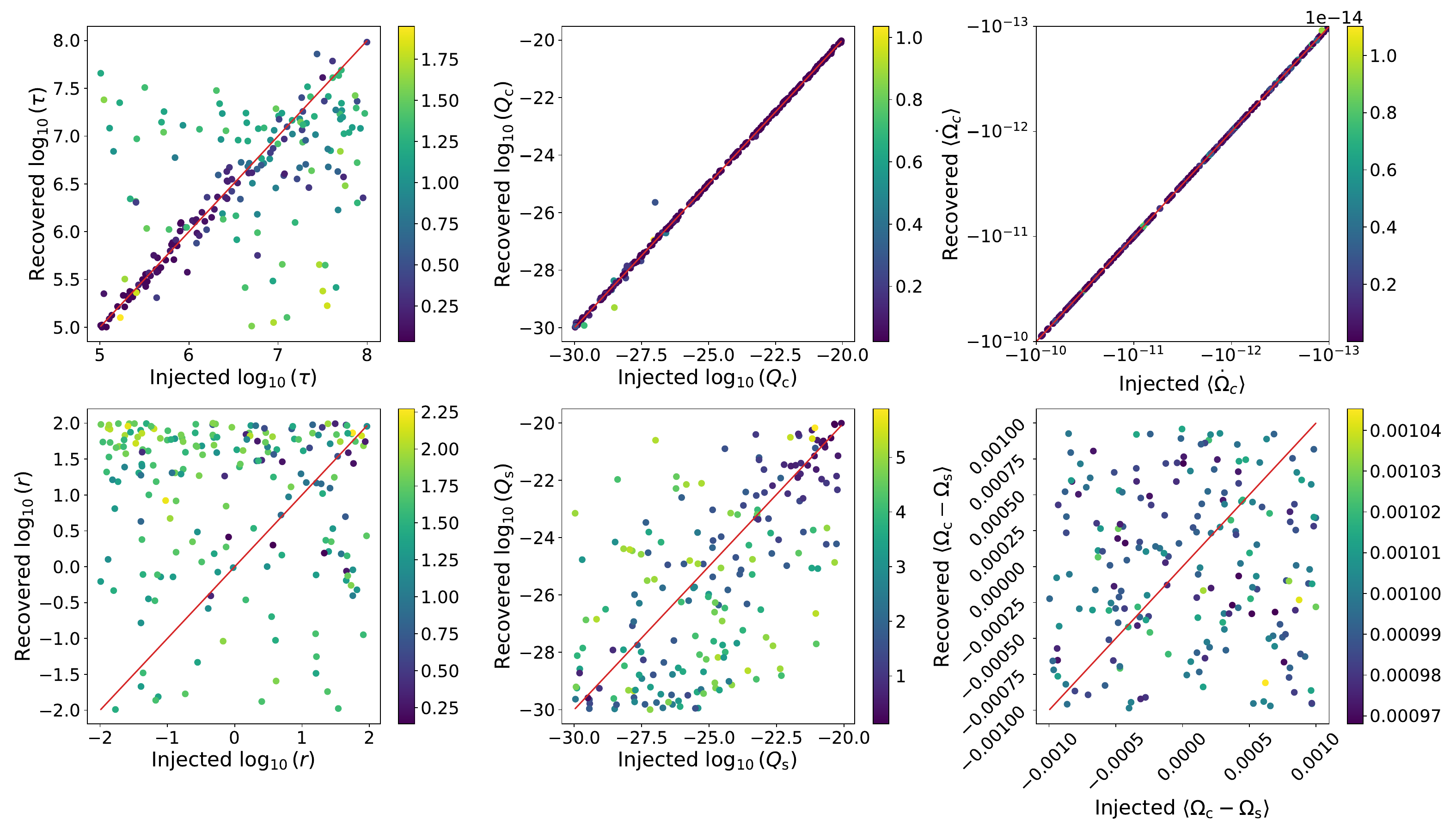}
    \caption{Test of the parameter estimation scheme on simulated frequency data showing the injected (horizontal axis) and recovered (vertical axis) $r, \tau, Q_{\rm{c}}, Q_{\rm{s}}, \langle \Omega_{\rm{c}}-\Omega_{\rm{s}} \rangle$ and $\langle \dot \Omega_{\rm{c}} \rangle$ values for 200 simulations. The injected parameters are those used to simulate the data and the recovered parameters are the peak of the posterior distribution obtained using the Kalman filter and nested sampler. The colour of each point can be compared to the colour bar next to its panel to get the width (interquartile range) of the marginalised posterior for that parameter.}
    \label{fig:simple_freq_test}
\end{figure*}

\begin{figure}
    \centering
\includegraphics[width=0.49\textwidth]{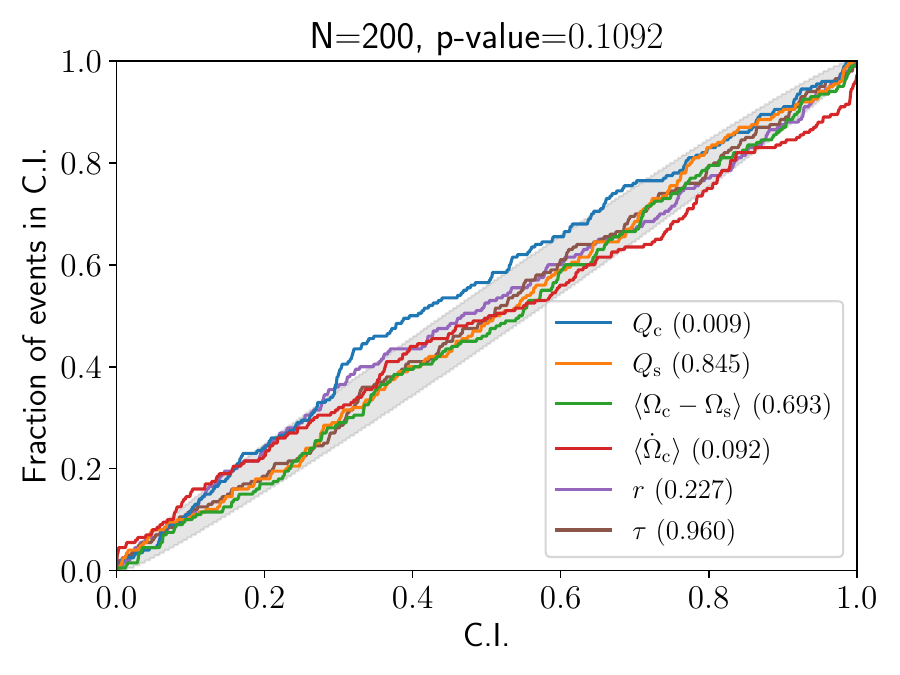}
    \caption{PP-plot for the 200 simulations in Fig. \ref{fig:simple_freq_test}, with the confidence interval on the horizontal axis and the proportion of simulations where the true values lie within that confidence interval on the vertical axis. The posterior distributions produced by the procedure in this paper are reliable if the curves stay close to the diagonal line. Specifically, if they stay within the shaded region then one can be 90\% confident that the posteriors are unbiased. For each parameter the legend lists the colour of the curve corresponding to it and $p$-values for the distribution of that parameter derived from a Kolmogorov-Smirnov test.}
    \label{fig:simple_freq_test_pp}
\end{figure}

Fig. \ref{fig:simple_freq_test} shows the results for 200 simulations with randomly chosen values for $r, \tau, Q_{\rm c}, Q_{\rm{s}}, \langle \Omega_{\rm{c}} - \Omega_{\rm{s}} \rangle, \langle \dot \Omega_{\rm{c}} \rangle$ and with $\Omega_{\rm{c}}(0) = 10~\rm{rad}~\rm{s}^{-1}$. The simulations are under idealised conditions (i.e.\ low noise, many samples) with $T_{\rm{obs}} = 1000$ days, $N_{\rm{obs}} = 1000$ and $R=10^{-30}~\rm{rad}^2\rm{s}^{-2}$. The six panels plot the recovered values for each of the six parameters against the injected parameters. The diagonal red lines mark where the recovered and injected parameters are equal, indicating a successful simulation. The colours indicate the width of the marginalised posterior for that parameter (width here meaning the inter-quartile range). 
The closeness of the points to the red line in the $\tau$, $Q_{\rm c}$ and $\langle \dot \Omega_{\rm{c}} \rangle$ panels means those parameters are generally well recovered, while the large vertical spread of the points in the $r$, $Q_{\rm{s}}$ and $\langle \Omega_{\rm{c}} - \Omega_{\rm{s}} \rangle$ panels means that they are usually harder to recover. This agrees with the identifiability analyses in Section \ref{ssec:identifiability} and Appendix \ref{app:identifiability of Qc, Qs and R}. 
Interestingly, small $\tau$ values tend to be better recovered, which is unsurprising since these correspond to strong damping.

As a further verification of the method, in Fig. \ref{fig:simple_freq_test_pp} we show a standard PP-plot (probability-probability plot) for the same 200 simulations as in Fig. \ref{fig:simple_freq_test}. The curves for each parameter remain close to the diagonal line, indicating that the posteriors give unbiased estimates. For more details on interpretation of PP-plots see \cite{Meyers2021b}. 

\subsection{Measurement noise}
\label{ssec:Effect of measurement noise on parameter estimation}

\begin{figure*}
    \centering
    \includegraphics[width=0.33\linewidth]{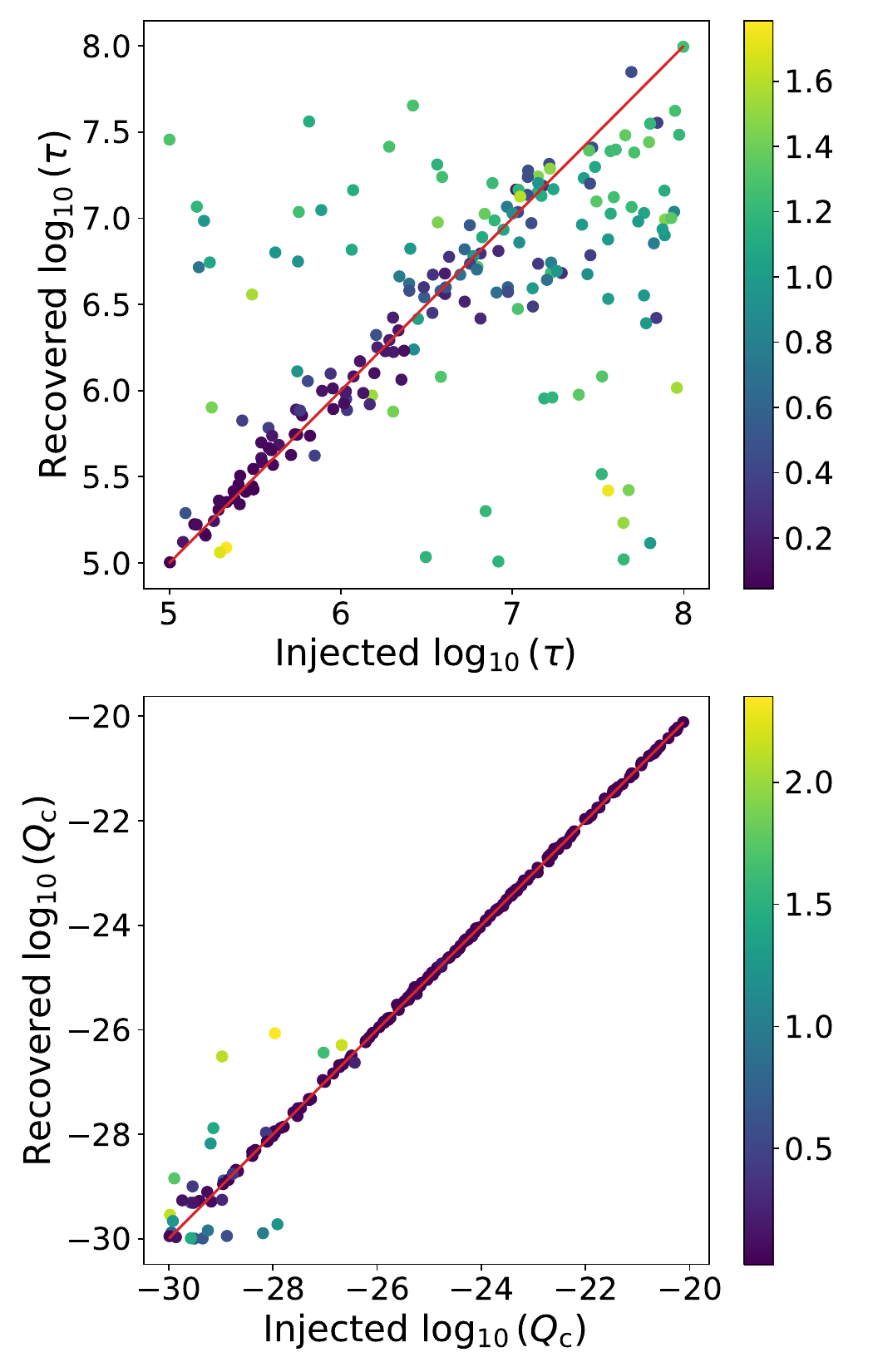}
    \includegraphics[width=0.33\linewidth]{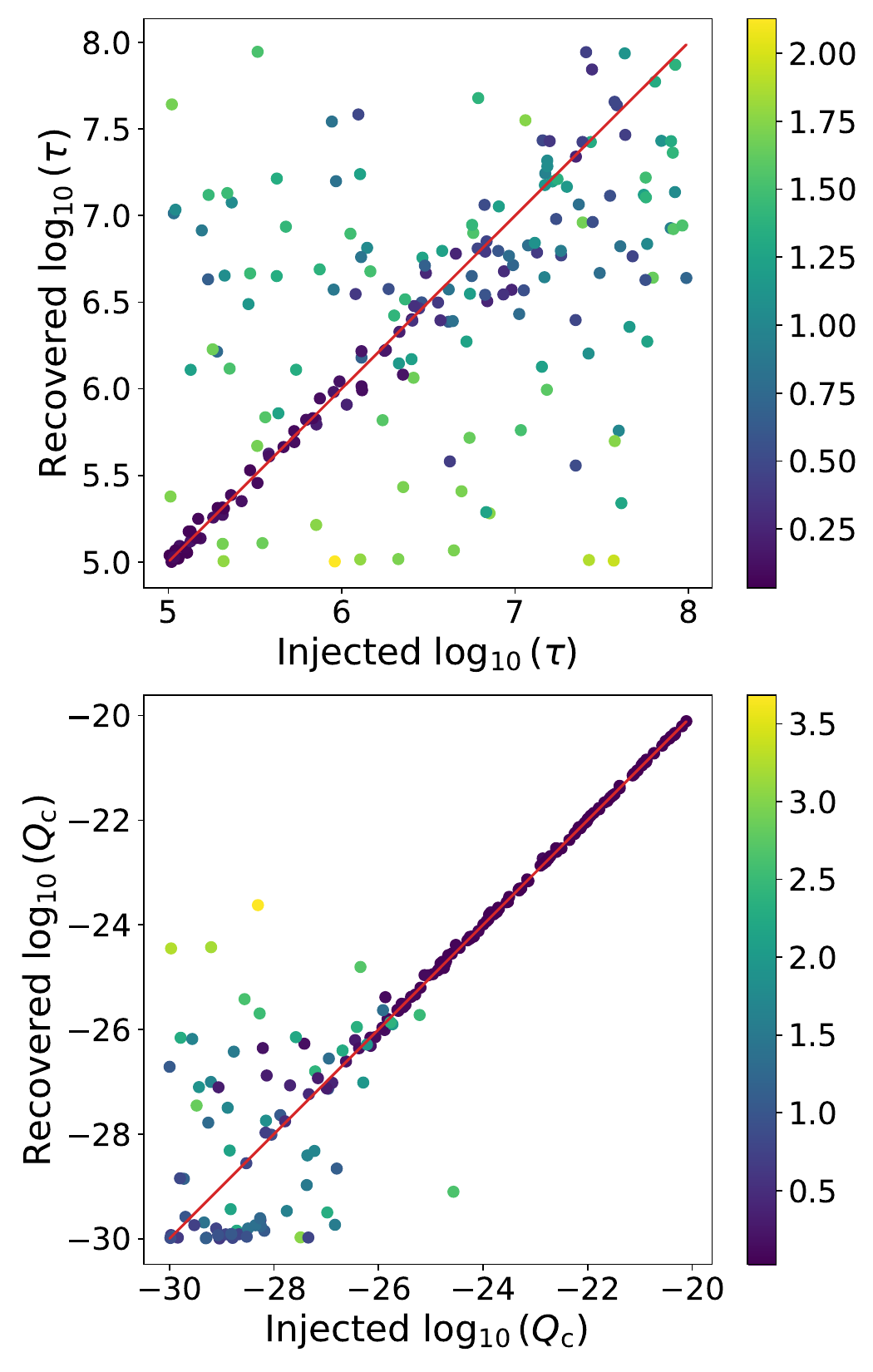}
    \includegraphics[width=0.33\linewidth]{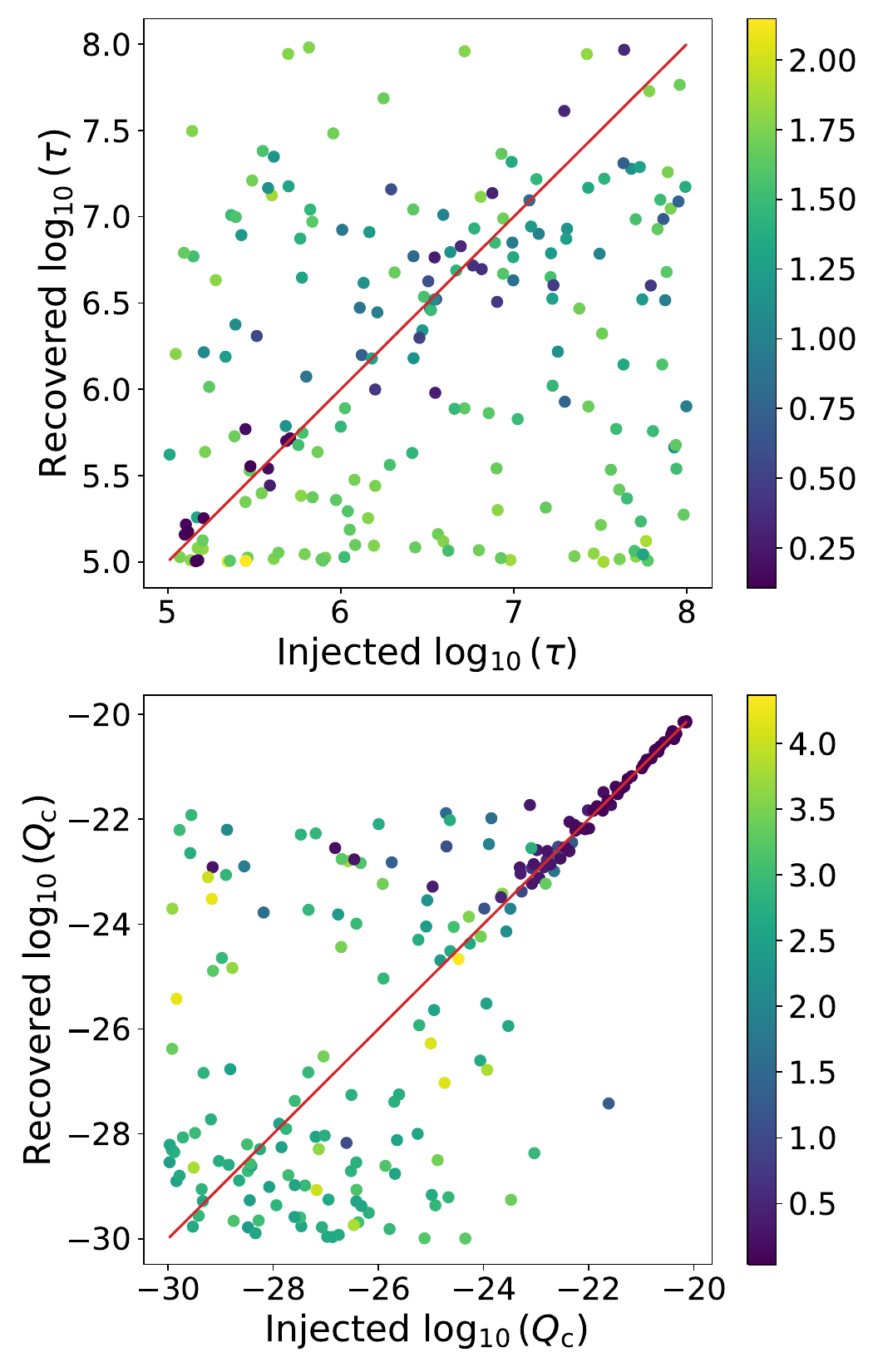}
    \caption{Test of the parameter estimation scheme on simulated frequency data showing the injected (horizontal axis) and recovered (vertical axis) $\tau$ and $Q_{\rm{c}}$ values (top and bottom rows respectively). Each column shows the results of 200 simulations with a different level of measurement error. Column 1 has $R = 10^{-24}~\rm{rad}^2\rm{s}^{-2}$, column 2 has $R = 10^{-20}~\rm{rad}^2\rm{s}^{-2}$, and column 3 has $R = 10^{-16}~\rm{rad}^2\rm{s}^{-2}$. 
    The colour of each point can be compared to the colour bar next to its panel to get the width (interquartile range) of the marginalised posterior for that parameter.}
    \label{fig:large_R_test}
\end{figure*}

In this appendix we test the Kalman filter's ability to recover parameters with different levels of measurement error.
Large measurement errors make it difficult for the Kalman filter to isolate the true random process, making the parameters harder to recover. 
To show the effect of large measurement noise we run three sets of 200 simulations, with $R=10^{-24}~\rm{rad}^2\rm{s}^{-2}$, $10^{-20}~\rm{rad}^2\rm{s}^{-2}$ and $10^{-16}~\rm{rad}^2\rm{s}^{-2}$. The results are shown in the three columns of Fig. \ref{fig:large_R_test}.
It is easiest to see the effect of varying $R$ by looking at $\tau$ and $Q_{\rm c}$ because they are the best recovered (except $\langle \dot \Omega_{\rm{c}} \rangle$ which is trivial).
We can see in each test that $Q_{\rm c}$ is poorly recovered below some critical value, when the measurement noise makes up most of the total noise. If $Q_{\rm c}$ is above the critical value, parameter estimation is usually successful.

\subsection{Impact of incorrect measurement uncertainties}
\label{ssec:effect of incorrect measurement uncertainties}

In this section we investigate the effect of feeding incorrect measurement uncertainties into the Kalman filter and introduce a modified parameter estimation algorithm to correct for the effect.

The frequency uncertainties for the real data in this paper are calculated from the TOA uncertainties provided in the UTMOST data release \citep{Lower2020}. 
Frequency uncertainties may be wrong, if the raw TOA errors or their conversion to frequency errors are incorrect.
The quoted TOA uncertainties depend on many factors such as dispersion measure \citep{2018CQGra..35m3001V}. 

\begin{figure}
    \centering
    \includegraphics[width=0.49\linewidth]{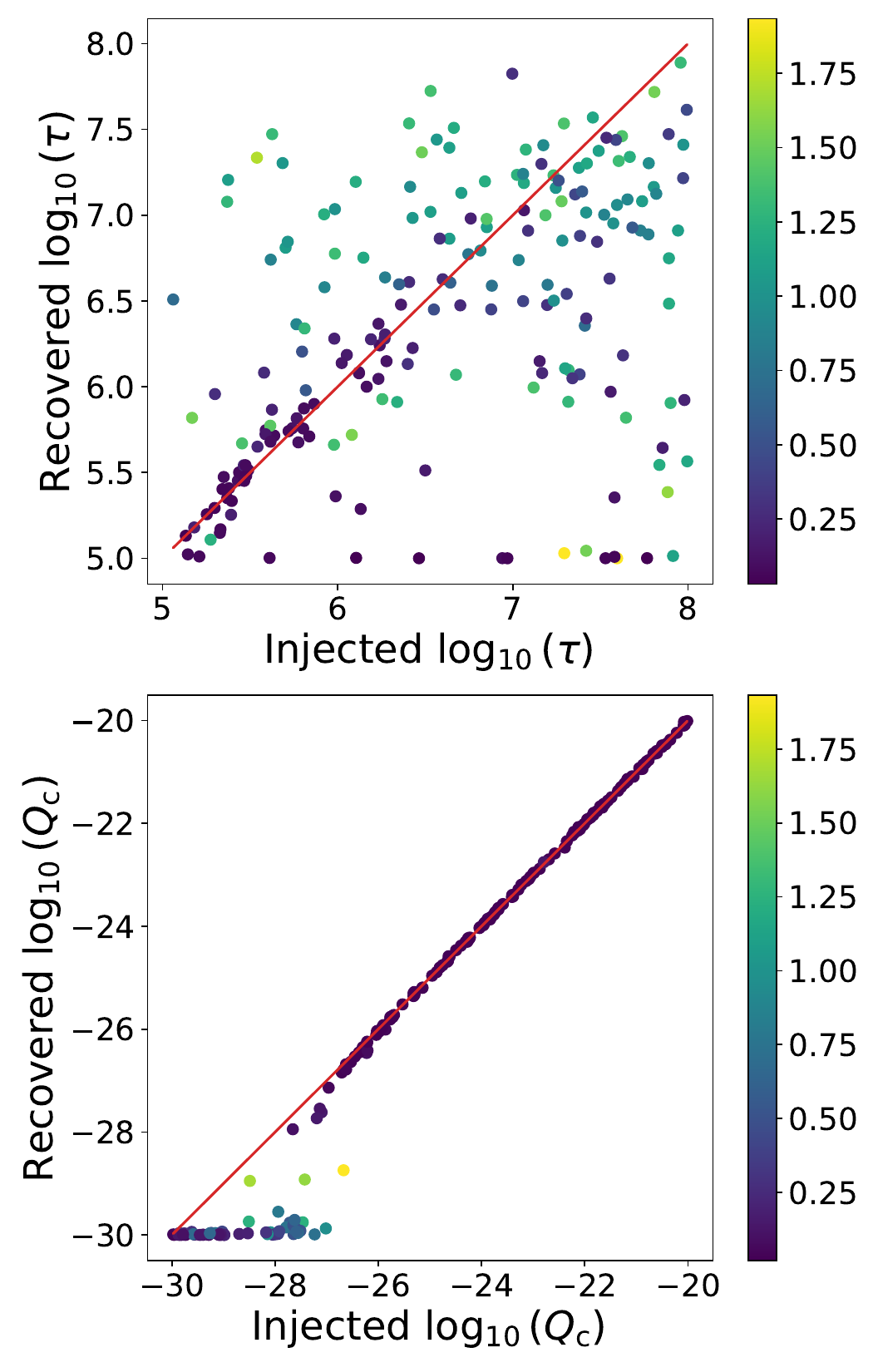}
    \includegraphics[width=0.49\linewidth]{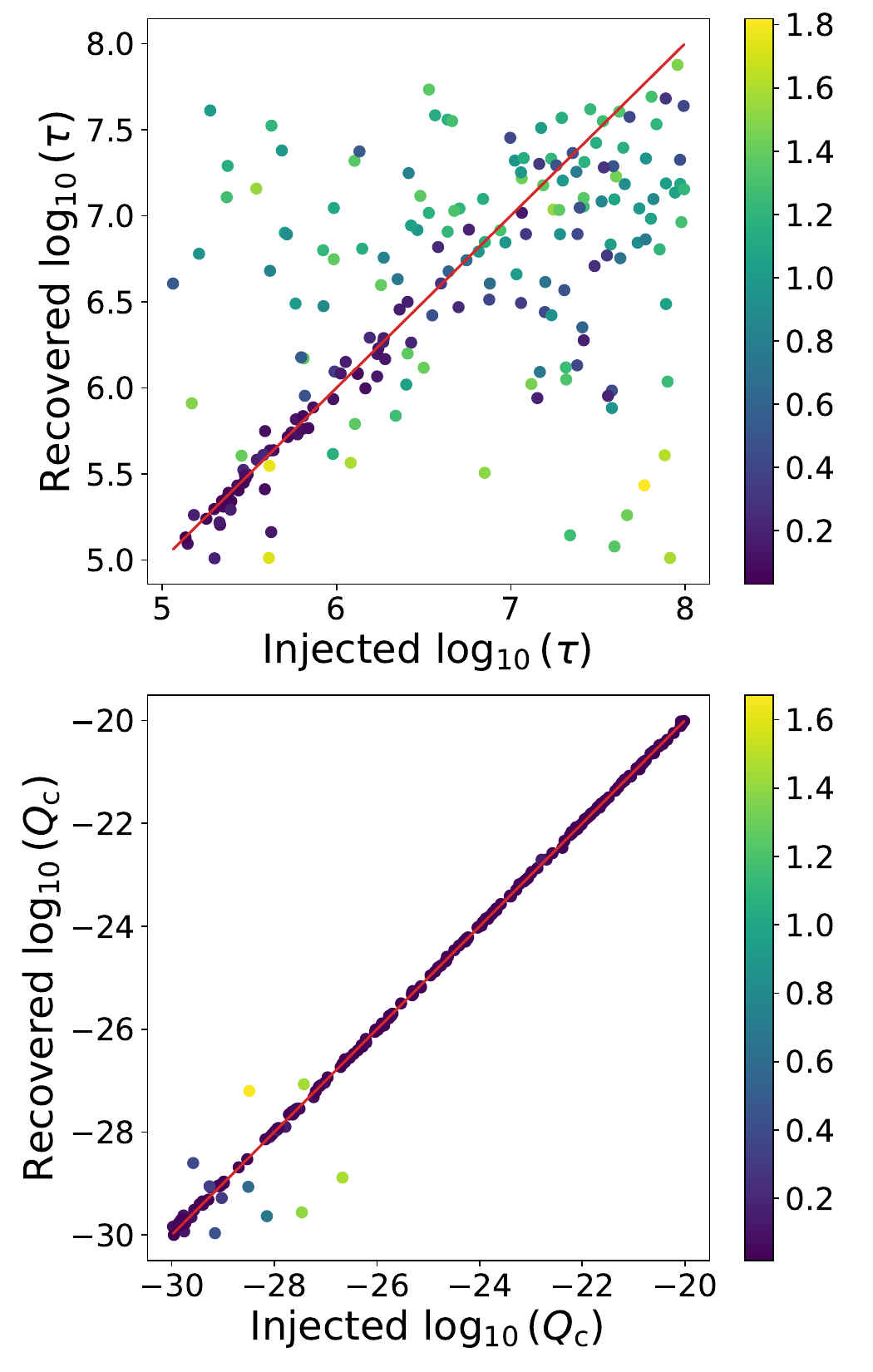}
    \caption{Test of the parameter estimation scheme on simulated frequency data showing the injected (horizontal axis) and recovered (vertical axis) $\tau$  and $Q_{\rm{c}}$ values (top and bottom rows respectively) for 200 simulations with $R_{\textrm{true}} = 10^{-27}~\rm{rad}^2\rm{s}^{-2}, R_{\textrm{KF}} = 10^{-23}~\rm{rad}^2\rm{s}^{-2}$. The colour of each point can be compared to the colour bar next to its panel to get the width (interquartile range) of the marginalised posterior for that parameter. The two plots in the left-hand column show the results with the normal parameter estimation algorithm. The plots in the right-hand column show the results on the same data supplemented by sampling of $R_{\rm KF}$.}
    \label{fig:EFAC_R_overest}
\end{figure}

\begin{figure}
    \centering
    \includegraphics[width=0.49\linewidth]{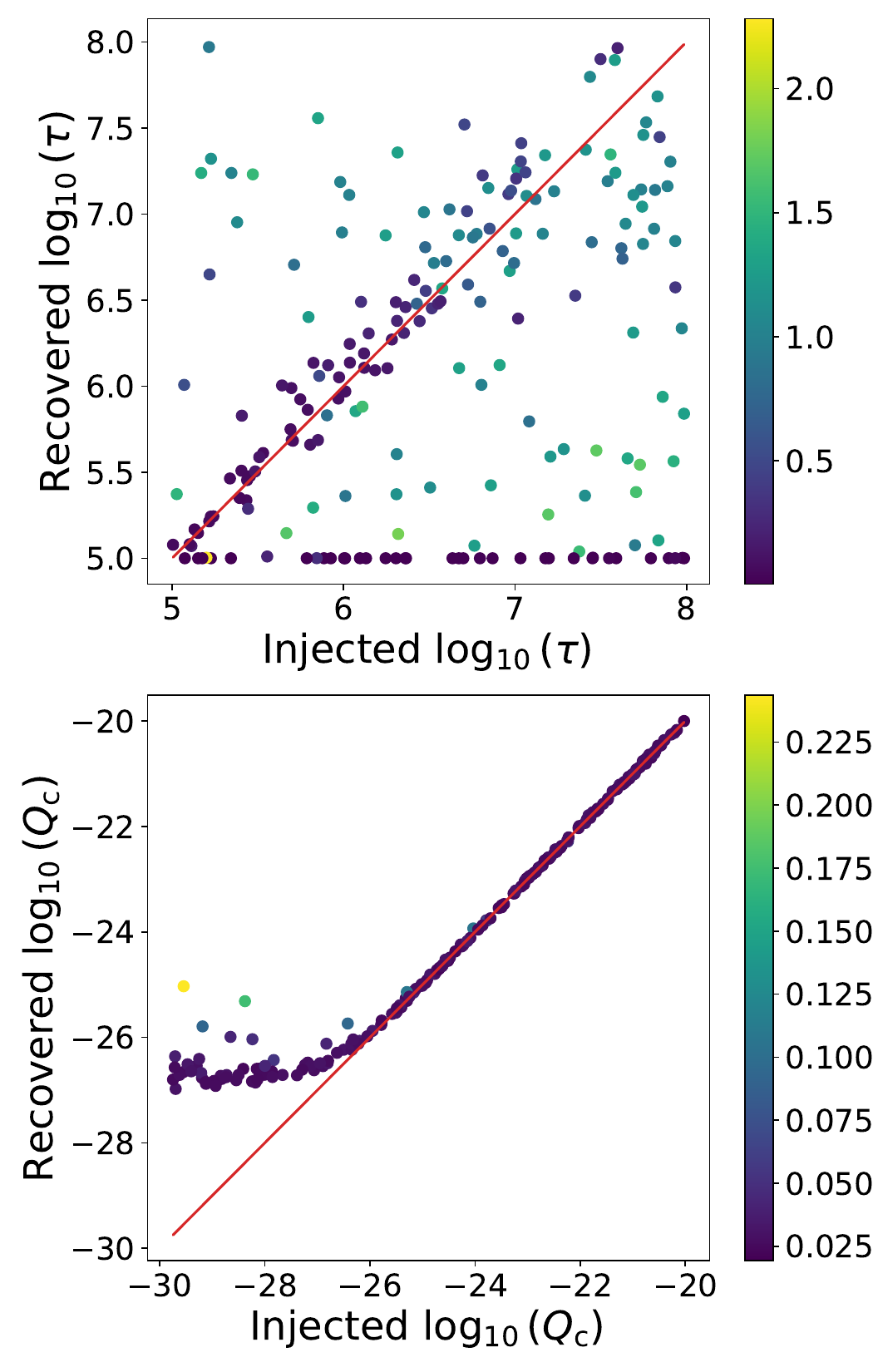}
    \includegraphics[width=0.49\linewidth]{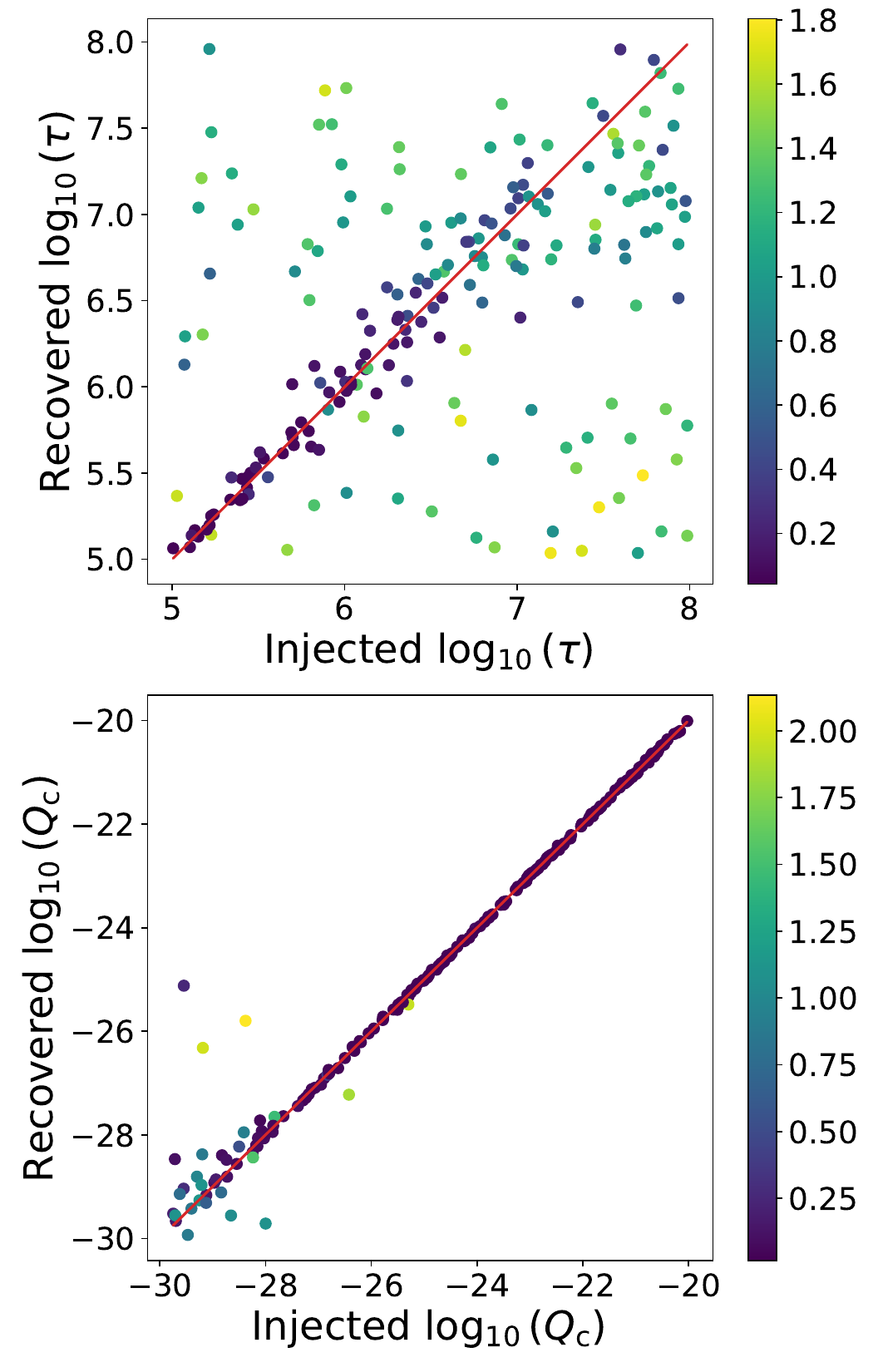}
    \caption{As for figure \ref{fig:EFAC_R_overest}, but with $R_{\rm true} = 10^{-23}~\rm{rad}^2\rm{s}^{-2}$ and $R_{\rm KF} = 10^{-27}~\rm{rad}^2\rm{s}^{-2}$.}
    \label{fig:EFAC_R_underest}
\end{figure}

To simulate the problem, we generate data with a measurement error variance $R_{\rm true}$ but use a different measurement error variance, $R_{\rm KF}$, in the Kalman filter. Results with $R_{\rm KF} = 10^{-23}~\rm{rad}^2\rm{s}^{-2}, R_{\rm true} = 10^{-27}~\rm{rad}^2\rm{s}^{-2}$ and $R_{\rm KF} = 10^{-27}~\rm{rad}^2\rm{s}^{-2}, R_{\rm true} = 10^{-23}~\rm{rad}^2\rm{s}^{-2}$ are shown in the left hand columns of Fig. \ref{fig:EFAC_R_overest} and Fig. \ref{fig:EFAC_R_underest} respectively. The results suggest that when $R_{\rm KF}$ is too large (column 1 of Fig. \ref{fig:EFAC_R_overest}) the $Q_{\rm c}$ values are underestimated and when $R_{\rm KF}$ is too small (column 1 of Fig. \ref{fig:EFAC_R_underest}) the $Q_{\rm c}$ values are overestimated. The effect is most pronounced for $Q_{\rm c} \lesssim 10^{-26}~\rm{rad}^2\rm{s}^{-3}$.

The Kalman filter separates measurement noise from the underlying random process guided by $R_{\rm KF}$. If it is guided to remove too much noise ($R_{\rm KF} \geq R_{\rm true}$), then some process noise is removed too, and the inferred strength of the process noise is weaker than it should be. The reverse is true for $R_{\rm KF} \leq R_{\rm true}$. If the measurement noise is not properly separated then $\tau$ is also difficult to recover.

We adjust for $R_{\rm KF}$ being input incorrectly by sampling $R$ as well as the six dynamical model parameters $r, \tau, Q_{\rm c}, Q_{\rm{s}}, \langle \Omega_{\rm{c}} - \Omega_{\rm{s}} \rangle, \langle \dot \Omega_{\rm{c}} \rangle$. The results are shown in the right columns of Fig. \ref{fig:EFAC_R_overest} and Fig. \ref{fig:EFAC_R_underest}. 
The recovered $Q_{\rm c}$ values are now close to their correct values and the spread on the recovered $\tau$ values is lower, which is an encouraging outcome. When analysing the real data in Section \ref{sec:results}, sampling $R$ makes little difference to the posterior, so the result is not displayed for brevity. We present this modified estimation method in this appendix in preparation for analysing more objects in the future.

\section{Identifiability of noise parameters}
\label{app:identifiability of Qc, Qs and R}

In this appendix we consider a simplified version of the parameter estimation problem to get a heuristic for the identifiability of the noise parameters $Q_{\rm{c}}$ and $Q_{\rm{s}}$, as we did in Section \ref{ssec:identifiability} for the other parameters. We aim to separate the measurement noise from $\xi_{\rm{c}}$ and $\xi_{\rm{s}}$ by exploiting their different behaviours over time. Specifically, $\xi_{\rm{c}}$ acts directly on the crust so it affects the frequency faster than $\xi_{\rm{s}}$, whose effect is delayed by the coupling time-scale. The measurement noise does not grow with time, nor does it evolve according to the equations of motion. The relative strengths of $\xi_{c}$, $\xi_{\rm{s}}$ and the measurement noise affect the size and shape of the observed random walk in $\Omega_{\rm c}$. By looking at the rate that the random walk in $\Omega_{\rm c}$ grows over different time-scales, the strengths of the three noise types can be estimated. In this section we set $N_{\rm c} = N_{\rm s} = 0$ to analyse just the random behaviour.

Let $\Omega_{\rm c}$ be observed at some instant and then again a time $\Delta t$ later, such that $\Omega_{\rm{c}}$ changes by an amount $\Delta \Omega_{\rm{c}}$. Because it is a random process, $\Delta \Omega_{\rm{c}}$ is drawn from a random distribution with a variance we can calculate as a function of $\Delta t$. By comparing the observed distribution of $\Delta \Omega_{\rm{c}}$ to the calculated distribution we can fit for the unknown parameters to infer $Q_{\rm{c}}$, $Q_{\rm{s}}$ and the size of the measurement noise.

The variance of a jump in $\Omega_{\rm c}$ over a time $\Delta t$ is given by equation (\ref{eq:process_noise_covariance_full}), viz.
\begin{align}
Q_{0,0}(\Delta t) &= \langle (\Omega_{\rm c}(t + \Delta t) - \Omega_{\rm c}(t))^2 \rangle\\
&= \frac{Q_{\rm c}\tau_{\rm c}^2+Q_{\rm s}\tau_{\rm s}^2}{(\tau_{\rm c}+\tau_{\rm s})^2}\Delta t\\
\nonumber&\quad+2\tau \taus \frac{Q_{\rm c}\tau_{\rm c} - Q_{\rm s}\tau_{\rm s}}{(\tau_{\rm c}+\tau_{\rm s})^2} \left(1 - e^{-\Delta t/\tau}\right)\\
\nonumber&\quad+\frac{\tau\taus^2}{2} \frac{Q_{\rm c} + Q_{\rm s}}{(\tau_{\rm c}+\tau_{\rm s})^2} \left(1 - e^{-2\Delta t/\tau}\right)\\
&= Q_{\rm c} T_{\rm c}(\Delta t) + Q_{\rm s} T_{\rm s}(\Delta t), \label{eq:omega jump var}
\end{align}
with
\begin{align}
T_{\rm{c}}(\Delta t) &= \frac{\tau_{\rm c}^2 \Delta t + 2 \tau \tau_{\rm c} \tau_{\rm s} \left(1 - e^{-\Delta t/\tau} \right) + \frac{\tau \tau_{\rm s}^2}{2} \left(1 - e^{-2\Delta t/\tau} \right)}{(\tau_{\rm c}+\tau_{\rm s})^2}\label{eq:Tc_formula}
\end{align}
and
\begin{align}
T_{\rm{s}}(\Delta t) &= \frac{\tau_{\rm s}^2 \Delta t - 2 \tau \tau_{\rm s}^2 \left(1 - e^{-\Delta t/\tau} \right) + \frac{\tau \tau_{\rm s}^2}{2} \left(1 - e^{-2\Delta t/\tau} \right)}{(\tau_{\rm c}+\tau_{\rm s})^2}. \label{eq:Ts_formula}
\end{align}
For $\Delta t \gg \tau$ one has 
\begin{align}
\label{eq:big var}
Q_{0,0} \approx \frac{Q_{\rm c}\tauc^2 + Q_{\rm s}\taus^2}{(\tauc+\taus)^2}\Delta t
\end{align}
and for $\Delta t \ll \tau$ one has 
\begin{align}
Q_{0,0} \approx Q_{\rm c} \Delta t.
\end{align}
Hence the influence of $\xi_{\rm{s}}$ through $Q_{\rm{s}}$ on short time-scales is negligible. 

The variance for a step in the measurement can be calculated from (\ref{eq:omega jump var}). Denote the measurement at time $t_i$ by $Y_i$ and the measurement noise by $v_i$. Then one has
\begin{align}
Y_i=\Omega_{c,i} + v_i.
\end{align}
In this appendix we assume for simplicity that all the measurement errors have variance $R$ even though in reality different data points have different measurement uncertainties.
Then the change in $Y$ between $t_i$ and $t_j$ has variance 
\begin{align}
\textrm{var}(\Delta Y) &= \textrm{var}(Y_j - Y_i)\\
&= \textrm{var}[(\Omega_{c,j}+v_j)-(\Omega_{c,i}+v_i)]\\
&= Q_{\rm c} T_{\rm c}(\Delta t) + Q_{\rm s} T_{\rm s}(\Delta t) + 2R.
\end{align}
Let us define the auxiliary function
\begin{align}
Q(\Delta t) = Q_{\rm c} T_{\rm c}(\Delta t) + Q_{\rm s} T_{\rm s}(\Delta t) + 2R.
\end{align}
Then $Y_j-Y_i$ is drawn from a normal distribution with variance $Q_{i,j} = Q(\Delta t_{i,j})$, with $\Delta t_{i,j} = t_j - t_i$.
Calculating $\tilde{Q}_{i,j}=(Y_j-Y_i)^2$ gives a single-sample estimate of the variance function $Q_{i,j}$. 
The average of many $\tilde{Q}_{i,j}$'s with the same $\Delta t$ converges to $Q(\Delta t)$.
Let us calculate $\tilde{Q}_{i,j}=(Y_j-Y_i)^2$ for the one-step transitions, $t_1 \rightarrow t_2$, $t_2 \rightarrow t_3$, ..., $t_{n-1} \rightarrow t_n$, two step transitions, $t_1 \rightarrow t_3$, ..., $t_{n-2} \rightarrow t_n$, all the way to the last and longest transition $t_1 \rightarrow t_n$. This makes $n(n-1)/2$ transitions. 
Let the estimate $\tilde{Q}_{i,j}$ differ from the true variance of the jump distribution, $Q_{i,j}$, by $\epsilon_{i,j}$, viz.
\begin{align}
\tilde{Q}_{i,j}=Q_{i,j}+\epsilon_{i,j}.
\end{align}
Then the set of equations for all the transitions is
\begin{align}
\tilde{Q}_{1,2} &= Q_{\rm c} T_{\rm c}(\Delta t_{1,2}) + Q_{\rm s} T_{\rm s}(\Delta t_{1,2}) + 2R + \epsilon_{1,2} \label{eq:regression_1}\\
\tilde{Q}_{2,3} &= Q_{\rm c} T_{\rm c}(\Delta t_{2,3}) + Q_{\rm s} T_{\rm s}(\Delta t_{2,3}) + 2R + \epsilon_{2,3}\\
\nonumber \vdots\\
\tilde{Q}_{1,n} &= Q_{\rm c} T_{\rm c}(\Delta t_{1,n}) + Q_{\rm s} T_{\rm s}(\Delta t_{1,n}) + 2R + \epsilon_{1,n}.\label{eq:regression_n}
\end{align}
If we assume we know $\tau_{\rm c}$ and $\tau_{\rm s}$ then we can calculate $T_{\rm c}$ and $T_{\rm s}$ using (\ref{eq:Tc_formula}) and (\ref{eq:Ts_formula}). 
Equations (\ref{eq:regression_1})-(\ref{eq:regression_n}) are a regression problem for the dependent variable $\tilde{Q}$ in terms of independent variables $T_{\rm{c}}$ and $T_{\rm{s}}$ where we fit for $Q_{\rm c}$, $Q_{\rm s}$ and $R$.
We make the simplification that the $\epsilon$'s are all drawn from the same distribution (this is not related to the $R$'s all being the same). Standard methods for least squares problems \citep{Draper1998} yield
\begin{align}
\begin{pmatrix}
Q_{\rm c}\\
Q_{\rm s}\\
2R
\end{pmatrix}
&=
\begin{pmatrix}
\langle T_{\rm c}^2 \rangle & \langle T_{\rm c} T_{\rm s} \rangle & \langle T_{\rm c} \rangle\\
\langle T_{\rm c} T_{\rm s} \rangle & \langle T_{\rm s}^2 \rangle & \langle T_{\rm s} \rangle\\
\langle T_{\rm c} \rangle & \langle T_{\rm s} \rangle & 1
\end{pmatrix}^{-1}
\begin{pmatrix}
\langle T_{\rm c} \tilde{Q}\rangle\\
\langle T_{\rm s} \tilde{Q}\rangle\\
\langle \tilde{Q}\rangle
\end{pmatrix},
\end{align}
with
\begin{align}
&\textrm{var}(Q_{\rm c}) = \frac{\langle T_{\rm s}^2 \rangle - \langle T_{\rm s} \rangle^2}{D} \textrm{var}(\epsilon), \label{eq:Qc est error}\\
&\textrm{var}(Q_{\rm s}) = \frac{\langle T_{\rm c}^2 \rangle - \langle T_{\rm c} \rangle^2}{D} \textrm{var}(\epsilon), \label{eq:Qs est error}\\
&\textrm{var}(R) = \frac{1}{4}\frac{\langle T_{\rm c}^2 \rangle \langle T_{\rm s}^2 \rangle - \langle T_{\rm c} T_{\rm s} \rangle^2}{D} \textrm{var}(\epsilon) \label{eq:R est error},
\end{align}
where $D$ is the determinant of the matrix being inverted,
\begin{align}
D &= (\langle T_{\rm c}^2 \rangle - \langle T_{\rm c} \rangle^2) (\langle T_{\rm s}^2 \rangle - \langle T_{\rm s} \rangle^2) - (\langle T_{\rm c} T_{\rm s} \rangle - \langle T_{\rm c} \rangle \langle T_{\rm s}\rangle)^2.
\end{align}

In equations (\ref{eq:Qc est error}) and (\ref{eq:Qs est error}), $\textrm{var}(Q_{\rm{c}})$ has $T_{\rm s}^2$ in the numerator, whereas $Q_{\rm{s}}$ has $T_{\rm{c}}^2$. We note that $T_{\rm{c}}$ is larger than $T_{\rm{s}}$ for small time steps. There are more one-step transitions than two-step transitions, more two-step than three-step transitions and so on. Hence we expect $\textrm{var}(Q_{\rm{c}})$ to be smaller than $\textrm{var}(Q_{\rm{s}})$ and $Q_{\rm{c}}$ should be easier to estimate from the data. Indeed, we do find that $Q_{\rm s}$ is difficult to identify in simulations, as discussed in Section \ref{ssec:simulation under ideal conditions}. The above calculation justifies the empirical results logically.

The results of the simulations shown in Fig. \ref{fig:simple_freq_test} show the relative identifiabilities of the six model parameters including the two process noise parameters $Q_{\rm c}$ and $Q_{\rm{s}}$. 
The recovered $Q_{\rm c}$ points lie close to the red line, indicating that they are usually recovered accurately. The recovered $Q_{\rm{s}}$ points are scattered over a wider range and are not recovered as well as $Q_{\rm c}$.

\section{tempo2 fits to local frequencies}
\label{app:tempo2 fits to local frequencies}

In this appendix we apply the parameter estimation scheme to frequencies fitted by {\sc tempo2} to simulated TOAs. We confirm that the low-pass filtering action of local frequency fitting introduces modest biases into the estimated parameters, chiefly $Q_{\rm c}$.

To simulate pulsar TOAs, we generate $\Omega_{\rm{c}}$ and $\Omega_{\rm{s}}$ data as in Appendix \ref{app:parameter estimation on simulated frequencies} but append a differential equation for the crust phase $\phi_{\rm{c}}$ to get
\begin{align}
\frac{d\Omega_{\rm{c}}}{dt} &= -\frac{1}{\tau_{\rm{c}}}(\Omega_{\rm{c}}-\Omega_{\rm{s}}) + \frac{N_{\rm{c}}}{I_{\rm{c}}} + \frac{\xi_{\rm{c}}}{I_{\rm{c}}} \label{eq:app crust freq de}\\
\frac{d\Omega_{\rm{s}}}{dt} &= -\frac{1}{\tau_{\rm{s}}}(\Omega_{\rm{s}}-\Omega_{\rm{c}}) + \frac{N_{\rm{s}}}{I_{\rm{s}}} + \frac{\xi_{\rm{s}}}{I_{\rm{s}}} \label{eq:app core freq de}\\
\frac{d\phi_{\rm{c}}}{dt} &= \Omega_{\rm{c}} \label{eq:app crust phase de}.
\end{align}
A TOA occurs when the pulsar beam points towards the Earth. We assume the beam is attached rigidly to the crust, so we define $\phi_{\rm{c}}=0~(\rm{mod}~2\pi)$ at a TOA. 
We then choose a set of observation epochs $t_1, t_2, ..., t_n$, which we anticipate adjusting slightly to coincide with the nearest TOAs. 
We integrate (\ref{eq:app crust freq de})--(\ref{eq:app crust phase de}) from $t_1$ [with $\phi_{\rm{c}}(t_1)=0$] to $t_2$. We extrapolate linearly to find the next nearest instant $t_2' > t_2$, that gives $\phi_{\rm c}(t_2')=0$, noting that linear extrapolation is safe, because the random noise contributes negligibly over a single rotation. We append $t_2'$ to the sequence of TOAs and repeat up to $t_3$ and so on. 
To simulate TOA measurement errors, random Gaussian noise is added to the TOAs. 
We then take each triple of consecutive TOAs and their uncertainties (defined as the standard deviation of the added noise) and use {\sc tempo2} to fit a frequency and a frequency uncertainty. 

The parameter estimation algorithm is applied to the fitted and exact $\Omega_{\rm{c}}$ values generated by (\ref{eq:app crust freq de})--(\ref{eq:app crust phase de}). We compare the two sets of results to assess the effect of the fitting process.
As in Appendix \ref{ssec:effect of incorrect measurement uncertainties}, we sample the measurement error. In this section, the measurement uncertainties are different for each data point. We vary the measurement error by sampling over two parameters, $\alpha$ and $\beta$, which correspond to the parameters EFAC and EQUAD commonly used in pulsar timing programs \citep{Edwards_2006a, 2014MNRAS.437.3004L}. Given values of $\alpha$ and $\beta$, we change the measurement error variances from $R$ to $R'$ according to the rule, 
\begin{align}
R' = \alpha R + \beta.
\end{align}

\begin{figure}
\centering
\includegraphics[width=0.49\linewidth]{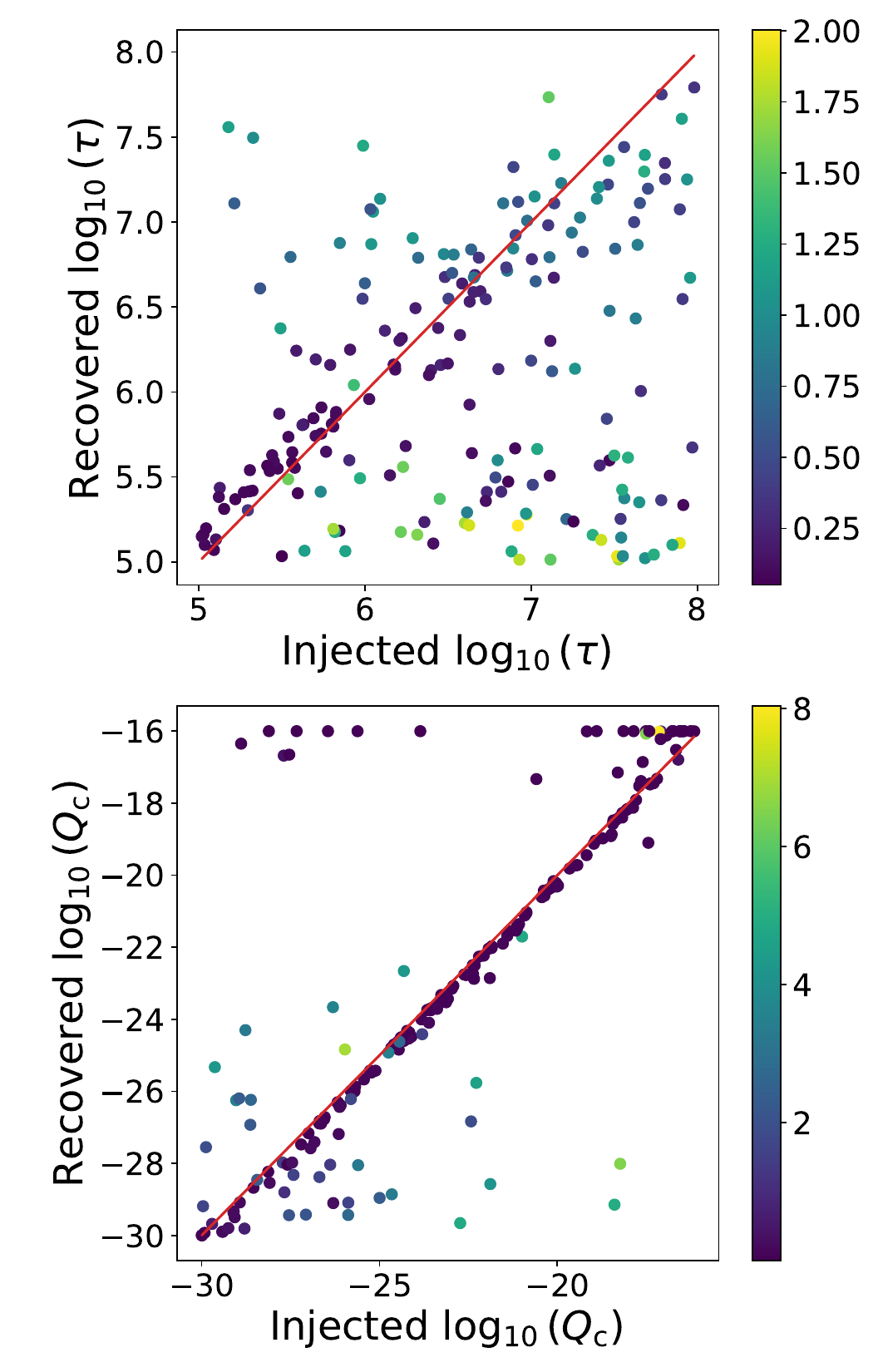}
\includegraphics[width=0.49\linewidth]{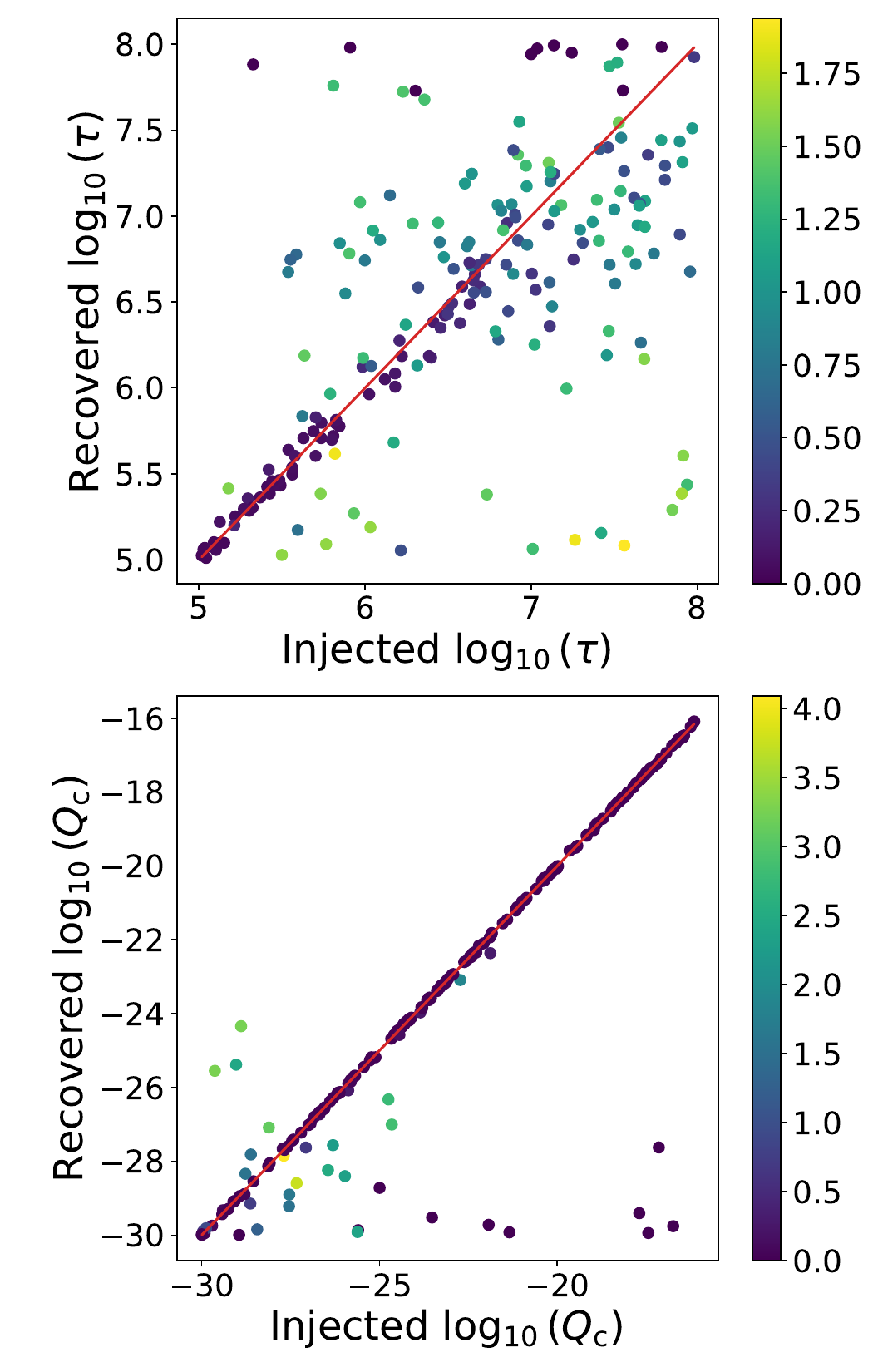}
\caption{Accuracy of {\sc tempo2} fits to local frequencies. Plots of the injected (horizontal axis) and recovered (vertical axis) $\tau$ and $Q_{\rm c}$ values (top and bottom rows respectively) for 200 simulations. The two plots in the left-hand column show the results when the method is applied to frequencies fitted to simulated TOAs with random spacings. The two plots in the right-hand column show the results when the method is applied to simulated frequencies without any {\sc tempo2} fitting. The colour of each point can be compared to the colour bar next to its panel to get the width (interquartile range) of the marginalised posterior for that parameter.}
\label{fig:tempo_test_random}
\end{figure}

We simulate ideal, data-rich conditions where $1000$ TOAs are spread over $1000$ days and the TOA uncertainties are all $10^{-16}~\rm{s}$.
The recovered $\tau$ and $Q_{\rm{c}}$ values for 200 simulations are shown in Fig. \ref{fig:tempo_test_random} for {\sc tempo2} fitted frequencies (left panels) and exact frequencies (right panels).
We find that $Q_{\rm{c}}$ and $\tau$ are estimated less accurately for fitted than for exact frequencies. $Q_{\rm{c}}$ is slightly underestimated for fitted frequencies; it is consistently just below the red diagonal. The $\tau$ values are also biased for the fitted frequencies, as can be seen from the many recovered $\tau$ values lying near the bottom of the prior range.

\begin{figure}
\centering
\includegraphics[width=0.49\linewidth]{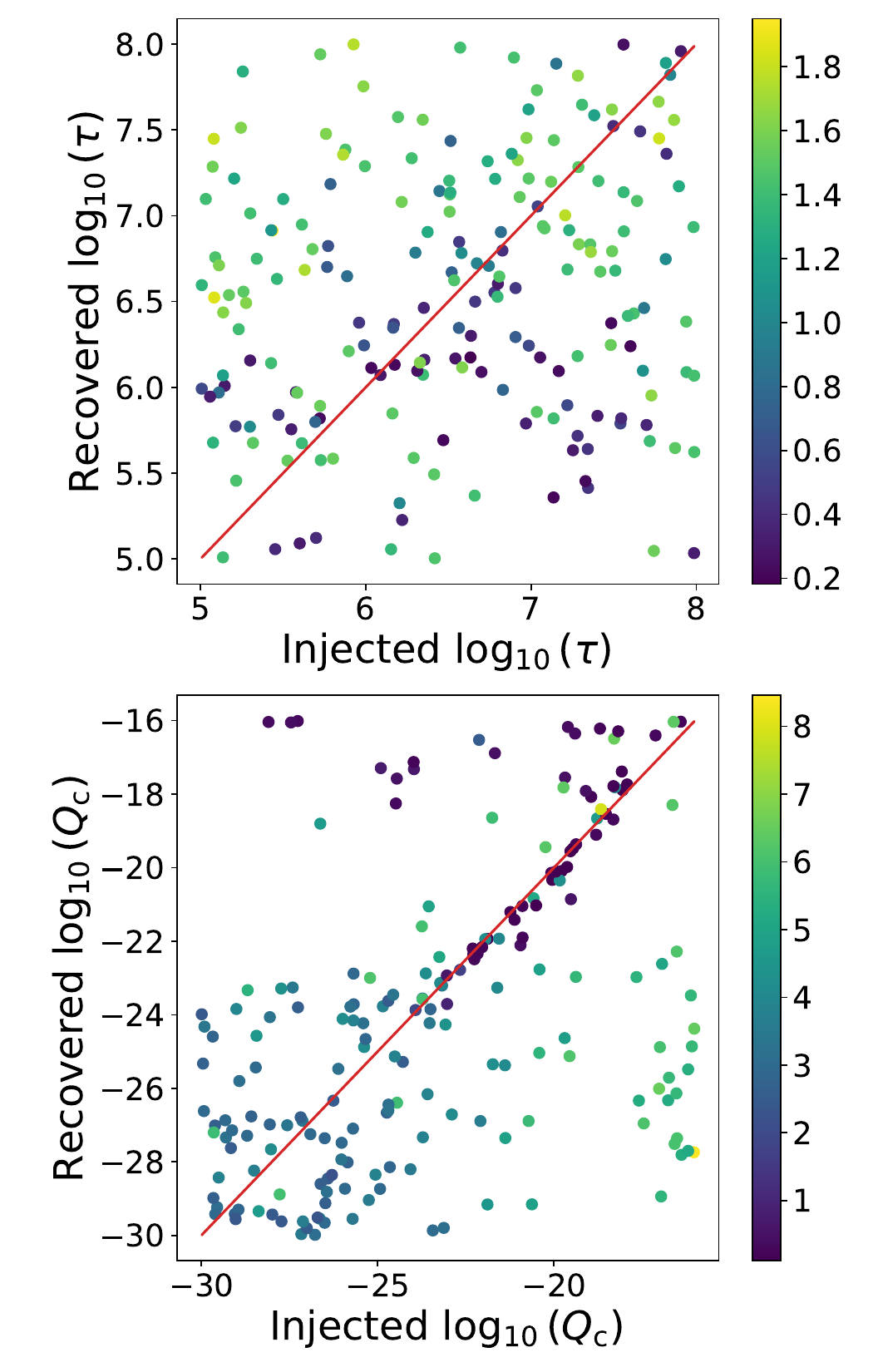}
\includegraphics[width=0.49\linewidth]{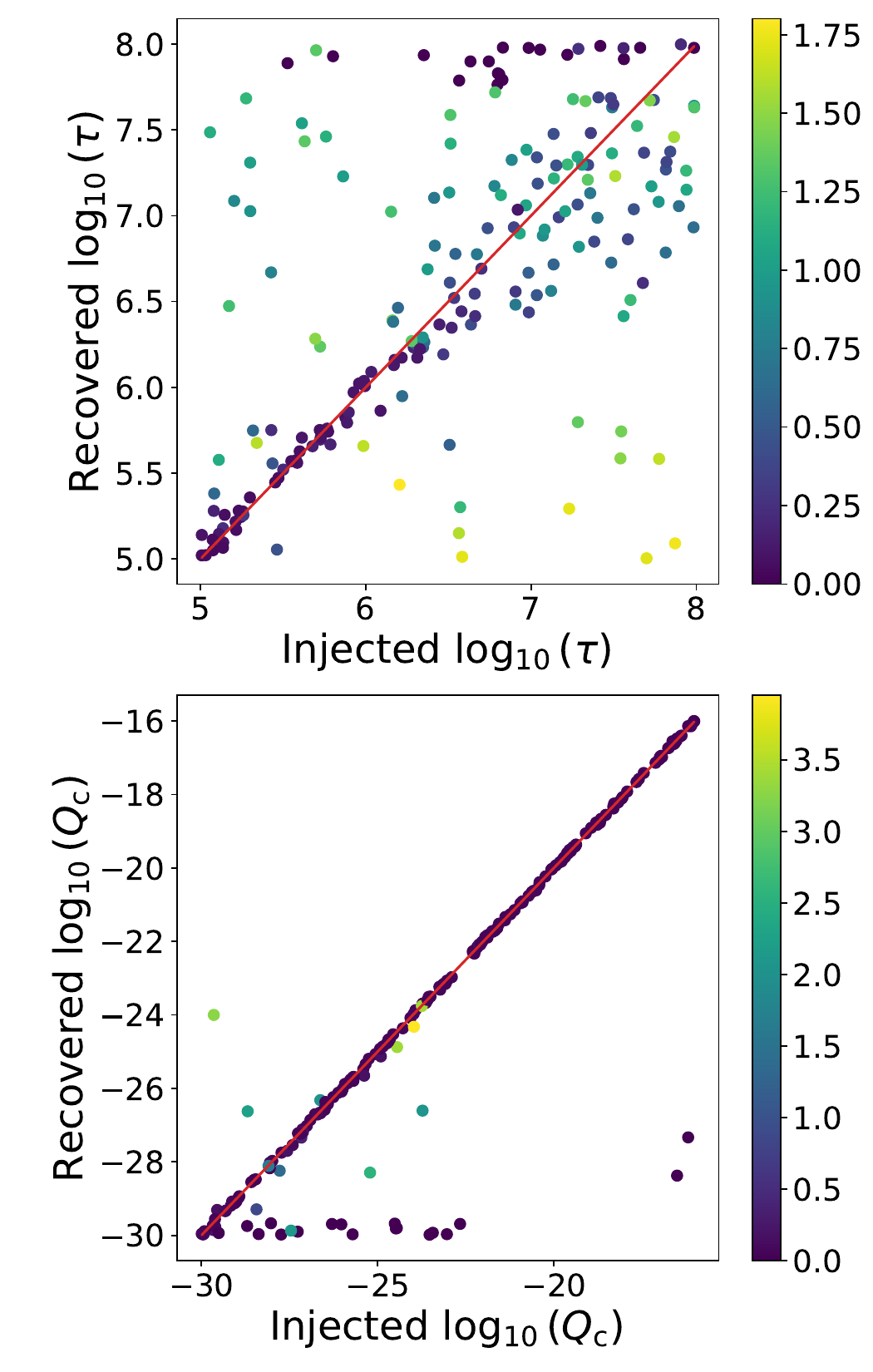}
\caption{Accuracy of {\sc tempo2} fits to local frequencies for synthetic data mimicking UTMOST observations of PSR J1359$-$6038. Plots of the injected (horizontal axis) and recovered (vertical axis) $\tau$ and $Q_{\rm c}$ values (top and bottom rows respectively) for 200 simulations. The two plots in the left-hand column show the results when the method is applied to frequencies fitted to simulated TOAs with the same spacing as the real data from PSR J1359$-$6038 used in Section $\ref{sec:results}$. The two plots in the right-hand column show the results when the method is applied to simulated frequencies with the same spacing and uncertainties as on the left, but without any {\sc tempo2} fitting. The colour of each point can be compared to the colour bar next to its panel to get the width (interquartile range) of the marginalised posterior for that parameter.}
\label{fig:tempo_test_cadence}
\end{figure}

\begin{figure}
\centering
\includegraphics[width=0.99\linewidth]{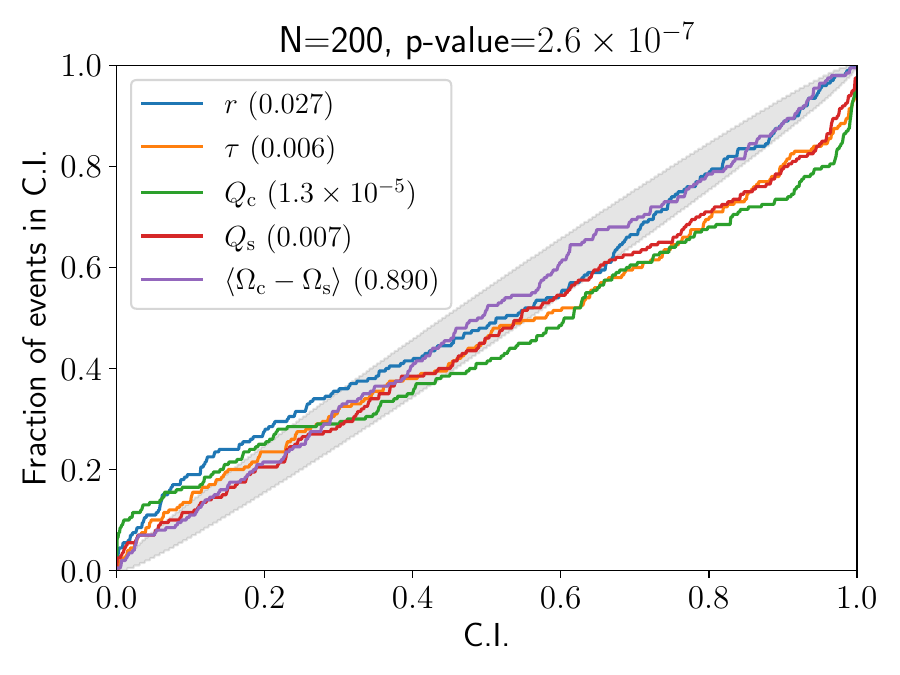}
\caption{PP-plot analogous to Fig. \ref{fig:simple_freq_test_pp} but for the 200 simulations in Fig. \ref{fig:tempo_test_cadence}.}
\label{fig:tempo_test_cadence_pp}
\end{figure}

We also simulate data with the same TOA spacings and measurement uncertainties as the real data in Fig. \ref{fig:freqs_J1359-6038} to quantify the effect of local fitting on parameter recovery under those conditions. The results are shown in Fig. \ref{fig:tempo_test_cadence} and Fig. \ref{fig:tempo_test_cadence_pp}. 
The plots of injected versus recovered parameters in Fig. \ref{fig:tempo_test_cadence} give an indication of how accurately the parameters are estimated.
In the top left panel, the recovered $\tau$ values have a large vertical spread. The $Q_{\rm{c}}$ values in the bottom left panel are also uncertain, especially for smaller injected $Q_{\rm{c}}$ values. 
The PP-plot in Fig. \ref{fig:tempo_test_cadence_pp} shows that the parameter curves deviate significantly from the diagonal.
However, the estimates do not seem to have a significant bias in either direction. 
So while the parameter estimates in Section \ref{sec:results} are uncertain due to the fitting procedure, they have no significant systematic bias.
The large spread of the recovered parameters is due in large part to the small number of frequency data points. More data would give a greater level of convergence to the true values as seen in Fig. \ref{fig:tempo_test_random}.

\section{Full posterior distribution for PSR J1359$-$6038}
\label{app:full posterior distribution for PSR J1359-6038}

\begin{figure*}
    \centering
    \includegraphics[width=0.9\textwidth]{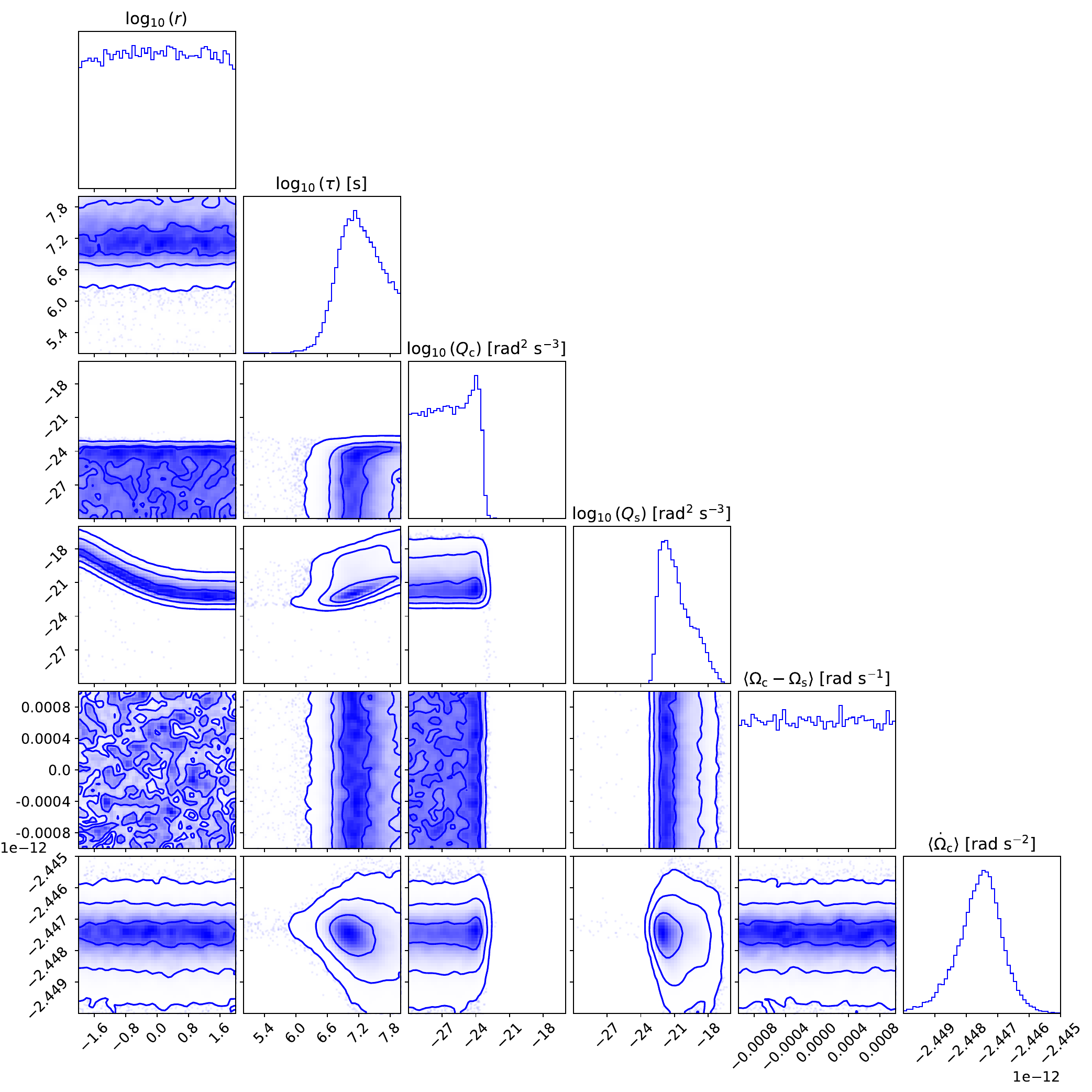}
    \caption{Six-dimensional joint posterior distribution for the two-component model for PSR J1359$-$6038. Marginalised posteriors for each parameter are shown as histograms on the diagonal. The lower half triangle of panels shows two-dimensional contour plots (with dark blue representing high probability) marginalised over all but two variables.}
    \label{fig:posteriors_J1359-6038}
\end{figure*}

In this appendix, for the sake of completeness, we present in Fig. \ref{fig:posteriors_J1359-6038} a visualisation of the posterior distribution $p(\boldsymbol \theta | \boldsymbol Y)$ for all six parameters $\theta = \{ r, \tau, Q_{\rm c}, Q_{\rm{s}}, \langle \Omega_{\rm{c}}-\Omega_{\rm{s}} \rangle, \langle \dot \Omega_{\rm{c}} \rangle \}$. The figure is formatted as a traditional corner plot: the panels with contours display $p(\boldsymbol \theta | \boldsymbol Y)$ marginalised over four out of six parameters, and the panels with histograms display $p(\boldsymbol \theta | \boldsymbol Y)$ marginalised over five out of six parameters. The distinctly peaked histograms in columns $2, 3, 4$ and $6$ are reproduced in Fig. \ref{fig:marginalised_posteriors_J1359-6038}. The flatter histograms in columns 1 and 5 correspond to parameters that cannot be inferred reliably from the data, as discussed in Section \ref{ssec:unidentifiable parameters}.

\section{One-component model}
\label{app:one-component model}

In this appendix, Bayesian model selection is performed to determine whether the two-component model outperforms a simpler one-component model when modelling PSR J1359$-$6038.

The model given by (\ref{eq:crust de}) and (\ref{eq:core de}) assumes there is a second component of the star, hidden from view, which couples to the crust, viz.\ the superfluid with angular velocity $\Omega_{\rm s}$. We could instead construct a simpler model where the pulsar is one rigid body described by the equation of motion
\begin{align}
I\frac{d\Omega}{dt} &= N + \xi(t), \label{eq:onecomp_torque}
\end{align}
with
\begin{align}
\langle \xi(t) \rangle &= 0\\
\langle \xi(t)\xi(t') \rangle &= \sigma^2 \delta(t-t') \label{eq:onecomp_noise},
\end{align}
where $\Omega$ is the angular velocity of the crust (to which the pulses are tied), $N$ is the constant spin-down torque, $I$ is the pulsar's moment of inertia, and $\xi$ is the stochastic torque responsible for timing noise modelled as a white noise process with strength $\sigma$.

We can make a Kalman filter for the one-component model and apply the same method of parameter estimation as in Section \ref{sec:kalman tracking and estimation}. For the one-component model, the update and measurement equations are still given by (\ref{eq:kalman update})-(\ref{eq:kalman measurement}) with
\begin{align}
F_i &= 1,\\
N_i &= \frac{N}{I} \Delta t_i,\\
Q_i &= \frac{\sigma^2}{I^2} \Delta t_i,\\
C &= 1,
\end{align}
and $\Delta t_i = t_i-t_{i-1}$.

We compare the two models' ability to explain the PSR J1359$-$6038 data by calculating the Bayesian evidence, $Z$, for each model given the data. We calculate the Bayesian evidence $Z_M$ for a model $M$ from the posterior distribution via the formula
\begin{align}
Z_M &= \int d\boldsymbol{\theta} p(\boldsymbol Y | \boldsymbol \theta, M) p(\boldsymbol \theta | M). \label{eq:evidence_integral}
\end{align}
The \texttt{dynesty} sampler generates samples from the posterior distribution to create the posterior plots and computes (\ref{eq:evidence_integral}) as a by-product.
Let $M_1$ and $M_2$ denote the one- and two-component models respectively.
We assert no prior preference for either model so set $p(M_1)=p(M_2)$, yielding the evidence ratio (Bayes factor)
\begin{align}
\frac{Z_1}{Z_2} &= \frac{p(M_1|\boldsymbol Y)}{p(M_2|\boldsymbol Y)}.
\label{eq:evidence_ratio}
\end{align}
Equation (\ref{eq:evidence_ratio}) is the relative preference for model 1 over model 2.

\begin{figure*}
    \centering
    \includegraphics[width=0.4\textwidth]{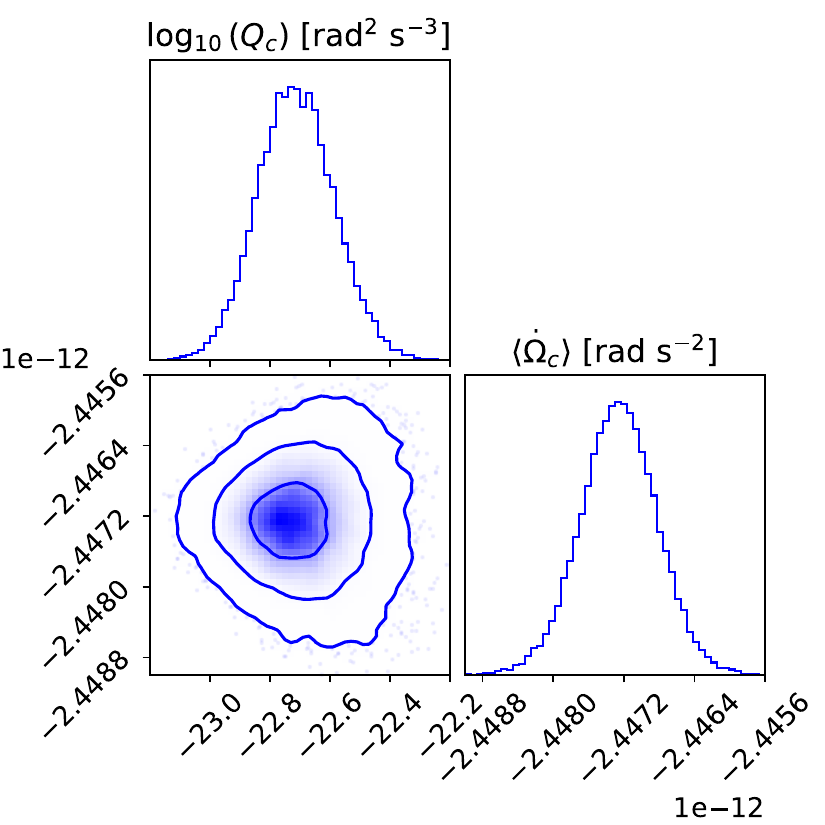}
    \caption{Two-dimensional joint posterior distribution for the one-component model for PSR J1359$-$6038. Marginalised posteriors for each parameter are shown as histograms on the diagonal. The bottom left panel shows the two-dimensional contour plot for the posterior.}
    \label{fig:posteriors_J1359-6038_onecomp}
\end{figure*}

Fig. \ref{fig:posteriors_J1359-6038_onecomp} shows the posterior distribution for the two parameters, $\langle \dot \Omega_{\rm{c}} \rangle = N/I$ and $Q_{\rm{c}} = \sigma_{\rm{c}}^2/I_{\rm{c}}^2$, obtained by applying the one-component model (\ref{eq:onecomp_torque})--(\ref{eq:onecomp_noise}) to the PSR J1359$-$6038 data from Section \ref{sec:data}. The key results of the parameter recovery are summarised in Table \ref{tab:onecomp_result_table}. The contour plot in Fig. \ref{fig:posteriors_J1359-6038_onecomp} shows no evidence for correlations between $\langle \dot \Omega_{\rm{c}} \rangle$ and $Q_{\rm{c}}$.

The logarithms of the Bayesian evidences calculated for the one- and two- component models are $\log_{10} Z_1 = 449.89$ and $\log_{10}Z_2 = 456.70$ respectively. The relative log Bayes Factor, $\log_{10} Z_2 - \log_{10} Z_1 = \Delta \log_{10} Z = 6.81$ categorically favours the two-component model. The uncertainties on $\log_{10} Z_1$ and $\log_{10} Z_2$ are $\sigma_1 = 0.017$ and $\sigma_2 = 0.017$ respectively, implying an approximate error for the log Bayes factor $\Delta \log_{10} Z$ of $(\sigma_1^2+\sigma_2^2)^{1/2} = 0.02$. 

\bgroup
\def\arraystretch{1.5}
\begin{table*}
    \centering
    \caption{Static parameters inferred by the Kalman tracker and nested sampler for the one-component model in appendix \ref{app:one-component model}, extracted from the two-dimensional joint posterior in Fig. \ref{fig:posteriors_J1359-6038_onecomp}. The fourth and fifth columns list two complementary measures of the dispersion in the posterior, viz.\ the full-width half-maximum (FWHM) and 90\% confidence intervals respectively.}
\begin{tabular}{c|c|c|c|c}
\hline
Parameter    & Units               & Peak value   & FWHM interval   & $90\%$ confidence interval\\
\hline
$\log_{10} Q_{\rm{c}}$ & $\rm{rad^2~s^{-3}}$ & $-22.73$ & $(-22.86,-22.56)$ & $(-22.92,-22.50)$\\   
$\langle \dot\Omega_{\rm c}\rangle$ & $\rm{rad~s^{-2}}$   & ~~$-2.4473 \times 10^{-12}$ & $(-2.4478 \times 10^{-12}, -2.4468 \times 10^{-12})$ & $(-2.4480 \times 10^{-12}, -2.4465 \times 10^{-12})$\\
\hline
\end{tabular}
\label{tab:onecomp_result_table}
\end{table*}
\egroup


\bsp	
\label{lastpage}
\end{document}